\documentclass[10pt,preprint]{emulateapj}
\usepackage{apjfonts}
\usepackage{latexsym,graphicx,rotating,amsmath, epsfig, natbib}
\renewcommand{\vec}{\boldsymbol}
\newcommand{\sol}{\odot}
\newcommand{\del}{\nabla}
\newcommand{\cross}{\vec{\times}}
\newcommand{\avg}{\bar}
\newcommand{\scrD}{\mathcal{D}}
\newcommand{\scrR}{\mathcal{R}}

\newcommand{\p}{\partial}

\newcommand{\nab}{\mbox{\boldmath $\nabla$}}
\newcommand{\rb}{\bar{\rho}}

\newcommand{\EE}{\cal E}
\newcommand{\FF}{\cal F}
\newcommand{\GG}{\cal G}

\newcommand{\VV}{\vec{v}}
\newcommand{\BB}{\vec{B}}

\newcommand{\Div}{\mbox{$\vec{\nabla} \cdot\,$}}
\newcommand{\advvm}{\left[ \langle v_r \rangle \frac{\p}{\p r}+\frac{ \langle v_{\theta} \rangle }{r}\frac{\p}{\p \theta}\right] }
\newcommand{\advvf}{\left[v_r'\frac{\p}{\p r}+\frac{v_{\theta}'}{r}\frac{\p}{\p \theta}+\frac{v_{\phi}'}{r\sin\theta}\frac{\p}{\p \phi}\right] }
\newcommand{\advbm}{\left[ \langle B_r \rangle \frac{\p}{\p r}+\frac{ \langle B_{\theta} \rangle }{r}\frac{\p}{\p \theta}\right] }
\newcommand{\advbf}{\left[B_r'\frac{\p}{\p r}+\frac{B_{\theta}'}{r}\frac{\p}{\p \theta}+\frac{B_{\phi}'}{r\sin\theta}\frac{\p}{\p \phi}\right] }

\newcommand\jfm{\rmfamily{J. Fluid Mech.}}
\newcommand\an{\rmfamily{Astron. Nachr.}}

\shorttitle{Persistent Wreaths of Magnetism}
\shortauthors{Brown, Browning, Brun, Miesch \& Toomre}

\begin{document}
\slugcomment{Published in ApJ}

  \received{May 27, 2009}
  \accepted{Dec 18, 2009}
  \published{Feb 11, 2010}

  \title{Persistent Magnetic Wreaths in a Rapidly Rotating Sun}
  \author{Benjamin P.\ Brown\altaffilmark{1}}
  \affil{JILA and Dept.\ Astrophysical \& Planetary Sciences,
    University of Colorado, Boulder, CO 80309-0440}
  \altaffiltext{1}{present address: Dept.\ Astronomy, University of Wisconsin,
    475 N.~Charter~St, Madison, WI 53706}
  \email{bpbrown@solarz.colorado.edu}
  \author{Matthew K.\ Browning}
  \affil{Canadian Institute for Theoretical Astrophysics, 
    University of Toronto, Toronto, ON M5S3H8 Canada}
  \author{Allan Sacha Brun}
  \affil{DSM/IRFU/SAp, CEA-Saclay and UMR AIM,
  CEA-CNRS-Universit\'e Paris 7, 91191 Gif-sur-Yvette, France}
  \author{Mark S.\ Miesch}
  \affil{High Altitude Observatory, NCAR, Boulder, CO 80307-3000}
  \and
  \author{Juri Toomre}
  \affil{JILA and Dept.\ Astrophysical \& Planetary Sciences, University of Colorado, Boulder, CO 80309-0440}

  \begin{abstract}
    When our Sun was young it rotated much more rapidly than
now. Observations of young, rapidly rotating stars indicate that many
possess substantial magnetic activity and strong axisymmetric magnetic
fields. We conduct simulations of dynamo action in rapidly rotating
suns with the 3-D MHD anelastic spherical harmonic (ASH) code to
explore the complex coupling between rotation, convection and
magnetism. Here we study dynamo action realized in the bulk of the
convection zone for a system rotating at three times the current solar
rotation rate.  We find that substantial organized global-scale
magnetic fields are achieved by dynamo action in this system. Striking
wreaths of magnetism are built in the midst of the convection zone,
coexisting with the turbulent convection. This is a surprise, for it
has been widely believed that such magnetic structures should be
disrupted by magnetic buoyancy or turbulent pumping.  Thus, many solar
dynamo theories have suggested that a tachocline of penetration and
shear at the base of the convection zone is a crucial ingredient for
organized dynamo action, whereas these simulations do not include such
tachoclines.  We examine how these persistent magnetic wreaths are
maintained by dynamo processes and explore whether a classical
mean-field $\alpha$-effect explains the regeneration of poloidal
field.  We find that the global-scale toroidal magnetic fields are
maintained by an $\Omega$-effect arising from the differential
rotation, while the global-scale poloidal fields arise from turbulent
correlations between the convective flows and magnetic fields.  These
correlations are not well represented by an $\alpha$-effect that is
based on the kinetic and magnetic helicities. 
  \end{abstract}


\section{Stellar Magnetism and Rotation}

Most stars are born rotating quite rapidly. They can arrive on the
main sequence with rotational velocities as high as 200 $\mathrm{km}\:\mathrm{s}^{-1}$
\citep{Bouvier_et_al_1997}.
Stars with convection zones at their surfaces, like the
Sun, slowly spin down as they shed angular momentum through their magnetized
stellar winds \citep[e.g.,][]{Weber&Davis_1967, Skumanich_1972,
  MacGregor&Brenner_1991}.  The time needed for significant spindown appears to be a
strong function of stellar mass \citep[e.g.,][]{Barnes_2003, West_et_al_2004}:
solar-mass stars slow less rapidly than somewhat less massive G and K-type
stars, but still appear to lose much of their angular momentum by the time
they are as old as the Sun.  Present day observations of the solar wind
likewise indicate that the current angular momentum flux from the Sun is a few
times 10$^{30}$ dyn~cm \citep[e.g.,][]{Pizzo_et_al_1983}, suggesting a time scale
for substantial angular momentum loss of a few billion years.  
Thus the Sun likely rotated significantly more rapidly in its youth than it does today.

\subsection{Rotation-Activity Relations} 
Rotation appears to be inextricably linked to stellar magnetic activity.
Observations indicate that in stars with extensive convective envelopes,
chromospheric and coronal activity -- which partly trace magnetic heating
of stellar atmospheres -- first rise with increasing rotation rate, then
eventually level off at a constant value for rotation rates above a
mass-dependent threshold velocity \citep[e.g.,][]{Noyes_et_al_1984a,
  Patten&Simon_1996, Delfosse_et_al_1998, Pizzolato_et_al_2003}.  
Activity may even decline somewhat in the most rapid rotators 
\citep[e.g.,][]{James_et_al_2000}.  Similar correspondence is
observed between rotation rate and estimates of the unsigned surface
magnetic flux \citep{Saar_1996, Saar_2001, Reiners_et_al_2009}.  This
``rotation-activity'' relationship is tightened when stellar rotation is
given in terms of the Rossby number $\mathrm{Ro} \sim P/\tau_c$, with $P$ the
rotation period and $\tau_c$ an estimate of the convective overturning time
\citep[e.g.,][]{Noyes_et_al_1984a}.  Expressed in this fashion, a common
rotation-activity correlation appears to span spectral types ranging from
late F to late M \citep[e.g.,][]{Mohanty&Basri_2003,
Pizzolato_et_al_2003, Reiners&Basri_2007}. 
Magnetic fields can likewise feed back upon stellar rotation by modifying
the rate at which angular momentum is lost through a stellar wind
\citep[e.g.,][]{Weber&Davis_1967, Matt&Pudritz_2008}.  Analyses of stellar spindown
as a function of age and mass have thus provided further constraints on
stellar magnetism and its connections to rotation. 
There are also indications that the period of the activity cycle
itself may depend on the stellar rotation rate
\citep[e.g.,][]{Saar&Brandenburg_1999}. 
Recent observations of solar-type stars may indicate
that even the topology of the global-scale fields changes with
rotation rate,  with the rapid rotators having substantial
global-scale toroidal magnetic fields at their surfaces \citep{Petit_et_al_2008}.
The overall picture that emerges from these observations is that rapid
rotation, as realized in the younger Sun and in a host of other stars, can
aid in the generation of strong magnetic fields and that
young stars tend to be rapidly rotating
and magnetically active, whereas older ones are slower and less active 
\citep[e.g.,][]{Barnes_2003, West_et_al_2004, West_et_al_2008}.  

A full theoretical understanding of the rotation-activity relationship, and
likewise of stellar spindown, has remained elusive.  Some aspects of these
phenomena probably depend upon the details of magnetic flux emergence,
chromospheric and coronal heating, and mass loss mechanisms -- but the
basic \emph{existence} of a rotation-activity relationship is widely
thought to reflect some underlying rotational dependence of the dynamo
process itself \citep[e.g.,][]{Knobloch_et_al_1981, Noyes_et_al_1984a,
Baliunas_et_al_1996}.

\subsection{Elements of Global Dynamo Action}
In stars like the Sun, the global-scale dynamo is generally thought to
be seated in the tachocline, an interface of shear between the
differentially rotating convection zone and the radiative interior
which is in solid body rotation
\citep[e.g.,][]{Parker_1993,Charbonneau&MacGregor_1997,Ossendrijver_2003}.
Helioseismology revealed the internal rotation profile of the Sun and
the presence of this important shear layer
\citep[e.g.,][]{Thompson_et_al_2003}.  The stably stratified
tachocline may provide a region for storing and amplifying
coherent tubes of magnetic field which may eventually rise to the
surface of the Sun as sunspots.  
Others have suggested that the latitudinal and radial gradients of
angular velocity in the bulk of the convection zone may be sufficient
for global dynamo action \citep[e.g.,][]{Dikpati&Charbonneau_1999,
Brandenburg_2005,  Guerrero&DalPino_2007}. 
However, it has generally been believed that
magnetic buoyancy instabilities may prevent fields from being strongly
amplified within the bulk of the convection zone itself
\citep{Parker_1975}.  In the now prevalent ``interface dynamo'' model,
solar magnetic fields are partly generated in the convection zone by
helical convection, then transported downward into the tachocline
where they are organized and amplified by the shear.  Ultimately the
fields may become unstable and rise to the surface.

Although the rotational dependence of this process is not well understood,
some guidance may come from mean-field dynamo theory.  In such theories,
the solar dynamo is often referred to as an ``$\alpha-\Omega$'' dynamo, with the
$\alpha$-effect characterizing the twisting of fields by helical convection
\citep[e.g.,][]{Moffatt_1978, Steenbeck_et_al_1966}, and the
$\Omega$-effect representing the shearing of poloidal fields by
differential rotation to form toroidal fields.  Both of these effects are,
in mean-field theory, sensitive to rotation: the $\alpha$-effect because
it is proportional to the kinetic helicity of the convective flows, which
sense the overall rotation rate, and the $\Omega$-effect because
more rapidly rotating stars are generally expected to have stronger
differential rotation.  But the detailed nature of these effects in
the solar dynamo and the appropriate scaling with rotation has been
very difficult to elucidate.

Simulations of the global-scale solar dynamo have generally affirmed
the view that the tachocline may play a central role in building the
globally-ordered magnetism in the Sun.  Recent three-dimensional (3D)
simulations of solar convection without a tachocline
at the base of the convection zone achieved dynamo action and produced
magnetic fields which were strongly dominated by fluctuating
components with little global-scale order \citep{Brun_et_al_2004}.
When a tachocline of penetration and shear was included, remarkable
global-scale magnetic structures were realized in the tachocline region, while
the convection zone remained dominated by fluctuating fields
\citep{Browning_et_al_2006}.  These simulations are making good progress 
toward clarifying the elements at work in the operation of the solar
global-scale dynamo, but for
other stars many questions remain.  In particular, observations of
large-scale magnetism in fully convective M-stars 
\citep{Donati_et_al_2006}, along with the persistence of a rotation-activity
correlation in such low-mass stars, hint that perhaps 
tachoclines may not be essential for the generation of global-scale
magnetic fields.  This view is partly borne out by simulations of
M-dwarfs under strong rotational constraints \citep{Browning_2008}, where
strong longitudinal mean fields were realized despite the
lack of either substantial differential rotation or a stable interior
and thus no classical tachocline.  
Major puzzles remain in the quest
to understand stellar magnetism and its scaling with stellar rotation.

\subsection{Convection and Dynamos in Rapidly Rotating Systems}
We began our study of rapidly rotating suns by carrying out a suite of
3D hydrodynamic simulations in full spherical shells that
explored the coupling of rotation and convection in these younger solar-type
stars \citep{Brown_et_al_2008}.  Those simulations studied the influence of rotation on the
patterns of convection and the nature of global-scale flows in such
stars.  The shearing flows of differential rotation generally grow
in amplitude with more rapid rotation, possessing rapid equators and
slower poles, while the meridional circulations weaken and break up
into multiple cells in radius and latitude.  More
rapid rotation can also substantially modify the patterns of
convection in a surprising fashion.
With more rapid rotation, localized states begin to appear in which
the convection at low latitudes is modulated in its strength with
longitude.  At the highest rotation rates, the convection can become
confined to active nests which propagate at distinct rates and persist for
long epochs. 

Motivated by these discoveries, we turn here to explorations
of the possible dynamo action achieved in a solar-type star rotating at
three times the current solar rate.  These 3D magnetohydrodynamic
(MHD) simulations span the convection zone alone, as the nature of
tachoclines in more rapidly rotating suns 
is at present unclear.  We find that a variety of dynamos can be
excited, including steady and oscillating states, and that dynamo action is
substantially easier to achieve at these faster rotation rates than in the solar
simulations.  Magnetism leads to strong feedbacks on the flows,
particularly modifying the differential rotation and its scaling with
the overall rotation rate $\Omega_0$.  The magnetic fields which form in these
dynamos have prominent global-scale organization within the convection zone, in contrast to
previous solar dynamo simulations \citep{Brun_et_al_2004, Browning_et_al_2006}.  

Quite strikingly, we find that
coherent global magnetic structures arise naturally in the midst of the
turbulent convection zone.  These wreath-like structures are regions of
strong longitudinal field $B_\phi$ organized loosely into tubes, with fields wandering in
and out of the surrounding convection.  These wreaths of magnetism differ
substantially from the idealized flux tubes supposed in many dynamo
theories, though they may be related to coherent structures
achieved in local simulations of dynamo action in shear flows
\citep{Cline_et_al_2003a, Vasil&Brummell_2008,Vasil&Brummell_2009}.  

Here we explore the nature of persistent magnetic wreaths realized in
a global simulation rotating at three times the solar rotation rate, and
discuss how they are maintained amidst turbulent convection.
In many of our other rapidly rotating suns, the dynamos become time
dependent and undergo semi-regular changes of global-scale polarity.
Those dynamos will be explored in an upcoming paper.  We additionally
find that magnetic wreaths survive in the presence of a model
tachocline, and those simulations will be reported on separately.

We outline in \S\ref{sec:ASH} the 3D MHD anelastic spherical shell
model and the parameter space explored by these simulations.  We then
examine in \S\S\ref{sec:steady_dynamo} and \ref{sec:wreaths} the
structure of magnetic fields found in our rapidly rotating dynamo at
three times the solar rate, which builds persistent global-scale
ordered fields in the form of wreaths in the midst of its convection zone. In
\S\ref{sec:dynamo_production} we examine how such global-scale fields
are created and maintained by dynamo processes. In \S\ref{sec:
mean-field} we explore whether a classical mean-field $\alpha$-effect
reproduces our observed production of poloidal field. We reflect on
our findings in \S\ref{sec:conclusions}.

\begin{deluxetable*}{ccccccccccccc}
   \tabletypesize{\footnotesize}
    \tablecolumns{13}
    \tablewidth{0pt}  
    \tablecaption{Parameters for Primary Simulations
    \label{table:sim_parameters}}
    \tablehead{\colhead{Case}  &  
      \colhead{$N_r,N_\theta,N_\phi$} &
      \colhead{Ra} &
      \colhead{Ta} &
      \colhead{Re} &
      \colhead{Re$'$} &
      \colhead{Rm} &
      \colhead{Rm$'$} &
      \colhead{Ro} &
      \colhead{Roc} &
      \colhead{$\nu$} &
      \colhead{$\eta$} &
      \colhead{$\Omega_0/\Omega_\sol$}
   }
   \startdata
    D3    & $96 \times 256 \times 512$ & 3.22$ \times 10^{  5}$ &     1.22$ \times 10^{  7}$ & 173 &  105 &   86 &   52 &    0.378 &    0.311 &     1.32 &     2.64 &  3 \\
    H3    & $96 \times 256 \times 512$ & 4.10$ \times 10^{  5}$ &     1.22$ \times 10^{  7}$ & 335 &  105 &  --- &  --- &    0.427 &    0.353 &     1.32 &     --- &  3 \\
    \enddata
 \tablecomments{Dynamo simulation at three times the solar rotation
    rate is case D3, and the hydrodynamic (non-magnetic) companion is H3.
        Both simulations have inner radius 
	$r_\mathrm{bot} = 5.0 \times 10^{10}$cm and outer radius of 
        $r_\mathrm{top} = 6.72 \times 10^{10}$cm, with 
	$L = (r_\mathrm{top}-r_\mathrm{bot}) = 1.72 \times 10^{10}$cm
	the thickness of the spherical shell.
	Evaluated at mid-depth are the
	Rayleigh number $\mathrm{Ra} = (-\partial \rho / \partial S)
	(\mathrm{d}\bar{S}/\mathrm{d}r) g L^4/\rho \nu \kappa$, 
	the Taylor number $\mathrm{Ta} = 4 \Omega_0^2 L^4 / \nu^2$, 
	the rms Reynolds number $\mathrm{Re}  = v_\mathrm{rms} L /\nu$ and
	fluctuating Reynolds number $\mathrm{Re}' = v_\mathrm{rms}' L /\nu$,
	the magnetic Reynolds number $\mathrm{Rm} = v_\mathrm{rms} L/\eta$ 
	and fluctuating magnetic Reynolds number 
        $\mathrm{Rm}' = v_\mathrm{rms}' L/\eta$, 
	the Rossby number $\mathrm{Ro} = \omega / 2 \Omega_0$ ,
	and the convective Rossby number 
	$\mathrm{Roc} = (\mathrm{Ra}/\mathrm{Ta} \, \mathrm{Pr})^{1/2}$.
	Here the fluctuating velocity $v'$ has the axisymmetric
        component removed: $v' = v - \langle v \rangle$, 
        with angle brackets denoting an average in longitude.
	For both simulations, the Prandtl number $\mathrm{Pr} = \nu / \kappa$ is 0.25 
	and in the dynamo simulation the magnetic Prandtl number
        $\mathrm{Pm}=\nu/\eta$ is 0.5.   
	The viscous and magnetic diffusivity, $\nu$ and $\eta$, are
	quoted at mid-depth (in units of $10^{12}~\mathrm{cm}^2\mathrm{s}^{-1}$).
        The rotation rate $\Omega_0$ of each reference frame is in multiples
        of the solar rate $\Omega_\sol=2.6 \times
        10^{-6}~\mathrm{rad}\:\mathrm{s}^{-1}$ or $414$ nHz.   
        The~viscous time scale at mid-depth $\tau_\nu = L^2/\nu$ is
        about $2600$~days for case D3 and the resistive time scale is
        about $1300$~days, while the rotation period is 9.3~days.
	}
\end{deluxetable*}

\section{Global Modelling Approach} 
\label{sec:ASH}

To study the coupling between rotation, magnetism and the
large-scale flows achieved in stellar convection zones, we must
employ a global model which 
simultaneously captures the spherical shell geometry and admits the
possibility of zonal jets and large eddy vortices, and of convective plumes
that may span the depth of the convection zone.  The solar convection
zone is intensely turbulent and microscopic values of viscosity and
magnetic and thermal diffusivities in the Sun are estimated to be very
small.  Numerical simulations cannot hope to resolve all scales of
motion present in real stellar convection and must instead strike a
compromise between resolving dynamics on small scales and capturing
the connectivity and geometry of the global scales. Here we focus on
the latter by studying a full spherical shell of convection.

\subsection{Anelastic MHD Formulation}
Our tool for exploring MHD stellar convection is the anelastic spherical
harmonic (ASH) code, which is described 
in detail in \cite{Clune_et_al_1999}.  The implementation of magnetism
is discussed in \cite{Brun_et_al_2004}.   
ASH solves the 3D MHD anelastic equations of motion in a
rotating spherical shell using the pseudo-spectral method and runs
efficiently on massively parallel architectures. 
We use the anelastic approximation to capture the effects of density
stratification without having to resolve sound waves which have short
periods (about 5 minutes) relative to the dynamical time scales of the
global scale convection (weeks to months) or possible cycles of
stellar activity (years to decades).  This criteria effectively filters
out the fast magneto-acoustic modes while retaining the slow modes and
Alfv\'en waves.
Under the anelastic approximation the thermodynamic fluctuating
variables are linearized about their spherically symmetric and
evolving mean state, with radially varying density $\bar{\rho}$, pressure $\bar{P}$,
temperature $\bar{T}$ and specific entropy $\bar{S}$.  The fluctuations
about this mean state are denoted as $\rho$, $P$, $T$ and $S$.  
In the reference frame of the star, rotating at average rotation rate $\Omega_0$, 
the resulting MHD equations are:
\begin{equation}
  \label{eq:div mass flux}
  \vec{\del} \cdot(\avg{\rho}\vec{v}) = 0\thinspace,
\end{equation}
\begin{equation}
  \label{eq:div B}
  \vec{\del} \cdot\vec{B} = 0\thinspace,
\end{equation}
\begin{equation}
  \label{eq:momentum}
  \begin{array}{c}
  \displaystyle \avg{\rho}\left[ \frac{\partial\vec{v}}{\partial t} +
    (\vec{v} \cdot \vec{\del})\vec{v} +
    2 \vec{\Omega}_0 \cross \vec{v} \right] 
  =  -\vec{\del} (\avg{P} + P) \\[3mm]
  \displaystyle + (\avg{\rho} + \rho) \vec{g}
                +\frac{1}{4 \pi} \left(\vec{\del} \cross \vec{B} \right) \cross \vec{B} 
                -\vec{\del} \cdot \scrD ,
  \end{array}
\end{equation}
\begin{equation}
  \label{eq:induction}
  \frac{\partial\vec{B}}{\partial t} = 
       \vec{\del} \cross (\vec{v} \cross \vec{B}) - \vec{\del} \cross (\eta \vec{\del} \cross \vec{B}), 
\end{equation}
\begin{equation}
  \label{eq:entropy}
  \begin{array}{c}
  \displaystyle 
  \avg{\rho}\avg{T}
  \left[\frac{\partial S}{\partial t} + \vec{v} \cdot \vec{\del}(\avg{S}+S)  \right]
  = \\[3mm]
  \displaystyle \vec{\del} \cdot \left[ \kappa_r \avg{\rho} c_p \vec{\del}(\avg{T}+T) 
                    +\kappa_0 \avg{\rho} \avg{T} \vec{\del} \avg{S}
		    +\kappa \avg{\rho} \avg{T} \vec{\del} S \right] \\[3mm]
  \displaystyle 
  + \frac{4 \pi \eta}{c^2} \vec{j}^2 
  + 2 \avg{\rho}\nu \left[e_{ij}e_{ij} - \frac{1}{3}(\vec{\del} \cdot \vec{v})^2\right],
  \end{array}
\end{equation}
where $\vec{v} = (v_r, v_\theta, v_\phi)$ is the local velocity
in the stellar reference frame,
$\vec{B}=(B_r, B_\theta, B_\phi)$ is the magnetic field, $\vec{j}$ is
the vector current density, 
$\vec{g}$ is the gravitational acceleration, 
$c_p$ is the specific heat at constant pressure, 
$\kappa_r$ is the radiative diffusivity and $\scrD$ is the viscous
stress tensor, given by
\begin{equation}
  \scrD_{ij} = -2 \avg{\rho} \nu \left[e_{ij} 
    - \frac{1}{3}(\vec{\del} \cdot \vec{v})\delta_{ij} \right],
\end{equation}
where $e_{ij}$ is the strain rate tensor.  Here $\nu$,
$\kappa$ and $\eta$ are the diffusivities for vorticity,
entropy and magnetic field.  We assume an ideal gas law
\begin{equation}
  \avg{P} = \scrR \avg{\rho} \avg{T},
\end{equation}
where $\scrR$ is the gas constant, and close this
set of equations using the linearized relations for the thermodynamic
fluctuations of
\begin{equation}
  \frac{\rho}{\avg{\rho}} = \frac{P}{\avg{P}} - \frac{T}{\avg{T}}
    =  \frac{P}{\gamma \avg{P}} - \frac{S}{c_p}.
\end{equation}
The mean state thermodynamic variables that vary with radius are
evolved with the fluctuations, thus allowing the convection to modify
the entropy gradients which drive it. 

The mass flux and the magnetic field are represented with a
toroidal-poloidal decomposition as
\begin{eqnarray}
  \avg{\rho}\vec{v} =& \vec{\del}\cross\vec{\del}\cross(W \hat{r}) + \vec{\del}\cross(Z\hat{r}), \\
            \vec{B} =& \vec{\del}\cross\vec{\del}\cross(\beta \hat{r}) + \vec{\del}\cross(\zeta\hat{r}),
\end{eqnarray}
with streamfunctions $W$ and $Z$ and magnetic potentials $\beta$ and $\zeta$.
This approach ensures that both quantities remain divergence-free to
machine precision throughout the simulation.  The velocity, magnetic
and thermodynamic variables are all expanded in spherical harmonics
for their horizontal structure and in Chebyshev polynomials for their
radial structure.  The solution is time evolved with a second-order
Adams-Bashforth/Crank-Nicolson technique.

ASH is a large-eddy simulation (LES) code, with subgrid-scale (SGS)
treatments for scales of motion which fall below the spatial resolution in our
simulations.  We treat these scales with effective eddy diffusivities,
$\nu$, $\kappa$ and $\eta$, which represent the transport of momentum,
entropy and magnetic field by unresolved motions in the simulations. 
These simulations are based on 
the hydrodynamic studies reported in \cite{Brown_et_al_2008}, and
as there $\nu$, $\kappa$ and $\eta$ are taken for simplicity 
as functions of radius alone and proportional to $\bar{\rho}^{-1/2}$.  
This adopted SGS variation, as in \cite{Brun_et_al_2004} and
\cite{Browning_et_al_2006}, yields lower diffusivities near the bottom of
the layer and thus higher Reynolds numbers. 
Acting on the mean entropy gradient is the eddy thermal diffusion $\kappa_0$
which is treated separately and occupies a narrow region in the upper
convection zone.  Its purpose is to transport entropy through the outer
surface where radial convective motions vanish.

The boundary conditions imposed at the top and bottom of the
convective unstable shell are:
\begin{enumerate}
  \item Impenetrable top and bottom: $v_r = 0\thinspace,$
  \item Stress-free top and bottom:
    \begin{equation}
       (\partial/\partial r)(v_\theta/r) =
      (\partial/\partial r)(v_\phi/r) = 0\thinspace, \nonumber
    \end{equation}
   \item Constant entropy gradient at top and bottom: 
    \begin{equation}
        \partial (S+ \bar{S})/\partial r = \mathrm{const},
    \end{equation}
   \item Match to external potential field at top:
    \begin{equation}
      B = \nabla \Phi  \quad \mathrm{and} \quad \nabla^2 \Phi = 0 {\big|}_{r=r_\mathrm{top}},\nonumber
    \end{equation}
  \item Perfect conductor at bottom:
    \begin{equation}
      B_r = (\partial/\partial r)(r B_\theta)=(\partial/\partial r)(r B_\phi)=0\thinspace.\nonumber
    \end{equation}
\end{enumerate}

\begin{figure*}
  \includegraphics[width=\linewidth]{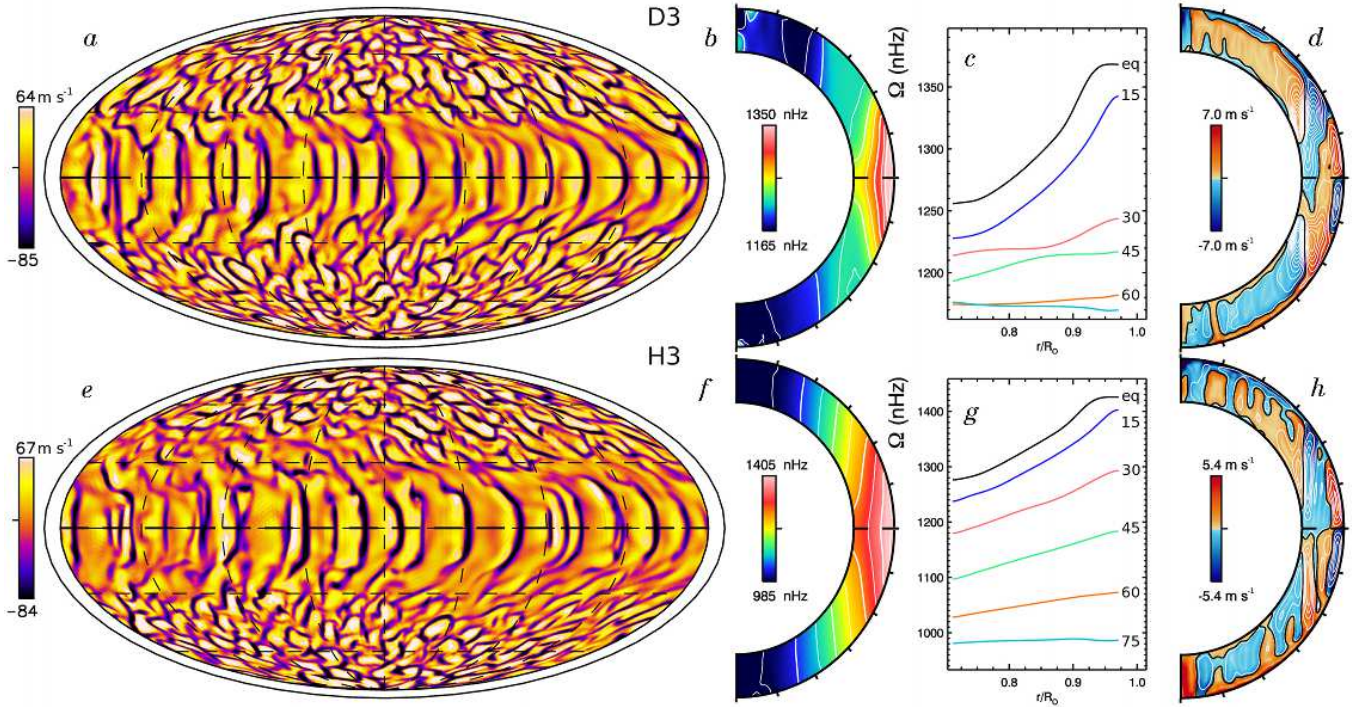}
  \caption{Convective structures and mean flows in cases D3 and H3. $(a)$~Radial
  velocity $v_r$ in dynamo case D3, shown in global Mollweide projection at
  $0.95R_\odot$, with upflows light and downflows dark.  Poles are at
  top and bottom and the equator is the thick dashed line.  The
  stellar surface at $R_\odot$ is indicated by the thin surrounding
  line.  $(b)$~Profiles of mean angular velocity $\Omega(r,\theta)$,
  accompanied in $(c)$~by radial cuts of $\Omega$ at selected
  latitudes.  A strong differential rotation is established by the
  convection. $(d)$~Profiles of meridional circulation, with sense of
  circulation indicated by color (red counter-clockwise, blue
  clockwise) and streamlines of mass flux overlaid.  
  $(e-h)$~Companion presentation of fields for hydrodynamic progenitor case H3.  The patterns of radial
  velocity are very similar in both cases.  The differential rotation
  is much stronger in the hydrodynamic case and the meridional
  circulations there are somewhat weaker, though their structure
  remains similar.
  \label{fig:case_D3_patterns}}
\end{figure*}

\subsection{Posing the Dynamo Problem}
Our simulations are a simplified picture of the vastly turbulent
stellar convection zones present in G-type stars.  We take solar
values for the input entropy flux, mass and radius, and explore
simulations of a star rotating at three times the current solar
rotation rate. We focus here on the bulk of the
convection zone, with our computational domain extending from
$0.72R_{\sol}$ to $0.97R_{\sol}$, thus spanning 172~Mm in radius.  The
total density contrast across the shell is about 25.  The
reference or mean state of our thermodynamic variables is derived from a
1D solar structure model \citep{Brun_et_al_2002} and is
continuously updated with the spherically symmetric components of the
thermodynamic fluctuations as the simulations proceed.  The reference
state in all of these simulations is similar to that shown in
\cite{Brown_et_al_2008}. We avoid regions near the stellar surface where hydrogen
ionization and radiative losses drive intense convection (like
granulation) on very small scales that we cannot resolve, and thus
position the upper boundary slightly below this 
region.  Our lower boundary is positioned near the base of the
convection zone, thus omitting the stably stratified
radiative interior and the shear layer at the base of the convection
zone known as the tachocline.   The fundamental characteristics of our
simulations and parameter definitions are summarized in
Table~\ref{table:sim_parameters}.

The dynamo simulation was initiated from a mature
hydrodynamic progenitor which had been evolved for more than 5000~days
and was well equilibrated.  The
progenitor case~H3 is very similar to case~G3 reported in
\cite{Brown_et_al_2008}, but here we chose a functional
form for the SGS entropy diffusion $\kappa_0$ that is more confined to the
upper 10\% of the convection zone; the unresolved flux here does 
not vary as much with rotation rate.
The effects of this change are subtle, resulting primarily in slightly
stronger latitudinal gradients of differential rotation and
temperature in the uppermost regions of the shell.  The patterns of
convection are very similar to those found in case~G3, though here
they are slightly more complex near the top
of the shell, and the Reynolds number remains high throughout the
convection zone.  Case~H3 possesses intricate convective patterns
and a solar-like differential rotation profile, with fast zonal flow at
the equator and slower flows at the poles.

\begin{figure*}
  \includegraphics[width=\linewidth]{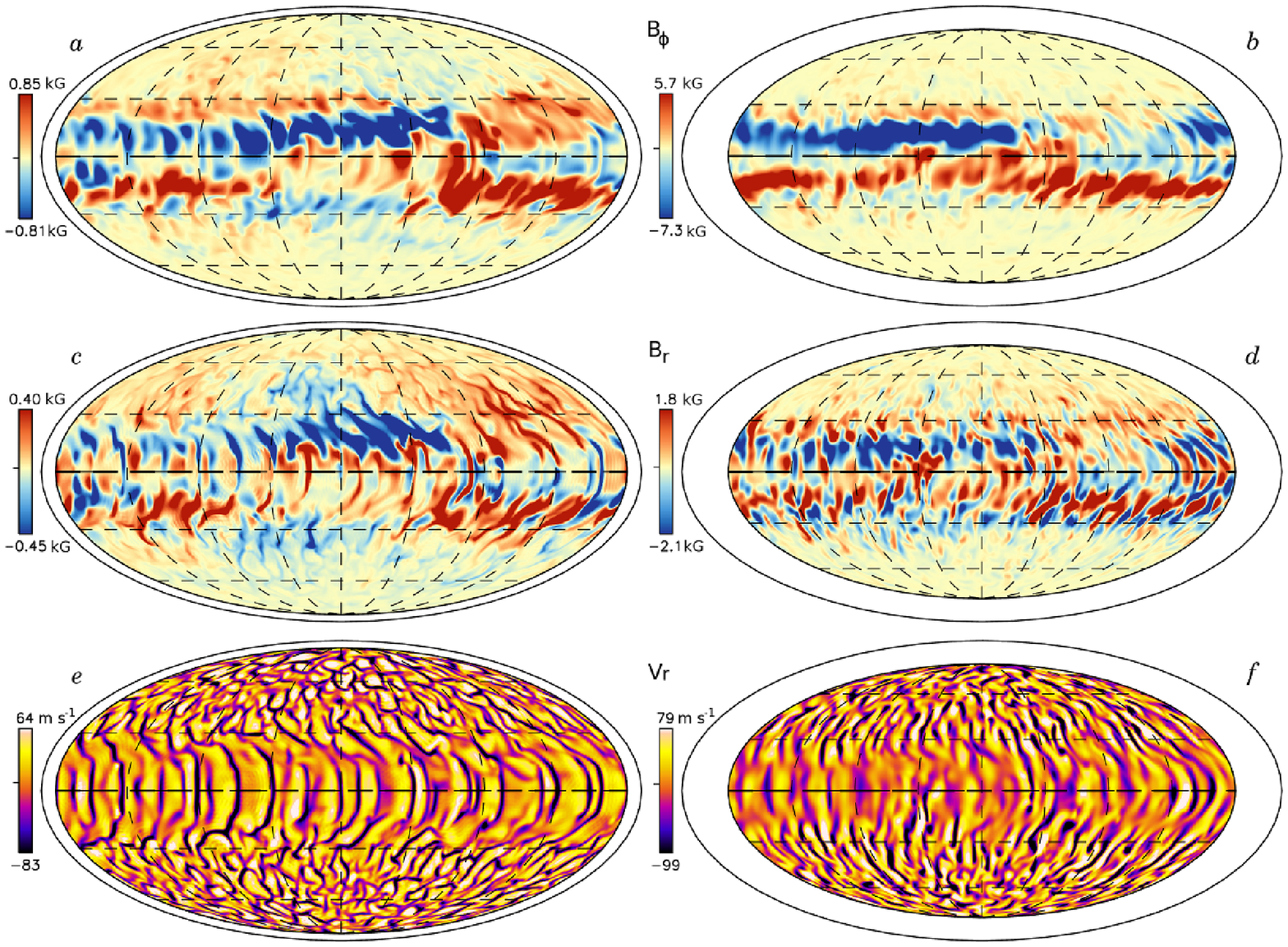}
  \caption{Magnetic wreaths and convective flows sampled at the same instant in
    case D3.   $(a)$~Longitudinal magnetic field $B_\phi$ near the top of the shell
    ($0.95R_\sol$) and ($b$)~at mid-depth ($0.85R_\sol$).  Strong
    flux structures with opposite polarity lie above and below the equator
    and span the convection zone.  ($c,d$)~Weaker radial magnetic
    field $B_r$ permeates and encircles each wreath. ($e,f$)~Strong convective upflows
    and downflows shown by $V_r$ pass through and around
    the wreaths.  The regions of strong magnetism tend to disrupt the
    convective flows while the strongest downflows serve to
    pump the wreaths to greater depths.
  \label{fig:case_D3_many_depths}}
\end{figure*}

To initiate our dynamo case, a small seed dipole magnetic field was
introduced and evolved via the induction equation.  The energy in the
magnetic fields is initially many orders of magnitude smaller than the energy
contained in the convective motions, but these fields are amplified by
shear and grow to become comparable in energy to the convective motions.

Stellar dynamo simulations are computationally intensive, requiring both
high resolutions to correctly represent the velocity fields and long
time evolution to capture the equilibrated dynamo behavior, which may
include cyclic variations on time scales of several years.  
The strong magnetic fields can produce rapidly moving Alfv\'en waves
which seriously restrict the Courant-Friedrichs-Lewy (CFL)
timestep limits in the upper portions
of the convection zone.    Case~D3, rotating three times faster than the
current Sun, has been evolved for over 7000 days (or over 2 million
timesteps).  We plan to report on a variety of other dynamo
cases, some at higher turbulence levels and rotation rates, in
subsequent papers.

This dynamo simulation was conducted at magnetic Prandtl number 
$\mathrm{Pm} = \nu/\eta =0.5$, a value significantly lower than
employed in our previous solar simulations.  In particular, \cite{Brun_et_al_2004}
explored $\mathrm{Pm} =2,2.5$ and $4$, and \cite{Browning_et_al_2006}
studied $\mathrm{Pm}=8$.  The high magnetic Prandtl numbers were required
in the solar simulations to reach sufficiently high magnetic Reynolds
numbers to drive sustained dynamo action.  In the simulations of
\cite{Brun_et_al_2004} only the simulations with $\mathrm{Pm} >2.5$
and $\mathrm{Rm}' \gtrsim 300$ achieved sustained dynamo action, where
$\mathrm{Rm}'$ is the fluctuating magnetic Reynolds number.
We are here able to use a lower magnetic Prandtl number for three
reasons.  Firstly, more rapid rotation tends to stabilize convection
and lower values of $\nu$ and $\eta$ are required to drive the
convection.  Once convective motions begin, they become quite
vigorous and the fluctuating velocities saturate at values comparable
to our solar cases.  Thus the Reynolds numbers achieved are
fairly large and we can achieve modestly high magnetic Reynolds
numbers even at low $\mathrm{Pm}$.  Secondly, the differential rotation becomes
substantially stronger with both more rapid rotation $\Omega_0$ and with
lower diffusivities $\nu$ and $\eta$.  This global-scale flow is an
important ingredient and reservoir of energy for these dynamos, and the
increase in its amplitude means that low $\mathrm{Pm}$ dynamos can
still achieve large magnetic Reynolds numbers based on this zonal flow. 
Lastly, the critical magnetic Reynolds number for dynamo action likely
decreases with increasing kinetic helicity
\citep[e.g.,][]{Leorat_et_al_1981}, and helicity generally increases with
rotation rate \citep[e.g.,][]{Kapyla_et_al_2009}.  Indeed there are
even suggestions that the presence of a mean shearing flow may lower
the critical magnetic Reynolds number \citep{Hughes&Proctor_2009}, and
the strong differential rotation present in these rapidly rotating
suns may serve to lower this threshold for dynamo action.
We find that the rapidly rotating flows considered here achieve dynamo
action at somewhat lower $\mathrm{Rm}$ than the models of
\cite{Brun_et_al_2004}, which rotated at the solar rate.

\section{Dynamos with Persistent Magnetic Wreaths}
\label{sec:steady_dynamo}

We here explore case~D3 which yields fairly persistent
wreaths of magnetism in its two hemispheres, though these do wax and
wane somewhat in strength once established.  Examining the properties
of this dynamo solution should help to provide a perspective for the greater
variations realized in our time-dependent dynamos which will be
discussed in a following paper.

\subsection{Patterns of Convection}
The complex and evolving convective structures in our dynamo cases are
substantially similar to the patterns of convection found in our
hydrodynamic simulations.  Our dynamo solution rotating at three times
the solar rate, case~D3, is presented in
Figure~\ref{fig:case_D3_patterns}, along with its hydrodynamic
progenitor, case~H3.  The radial velocities shown near the
top of the simulated domain (Figs.~\ref{fig:case_D3_patterns}$a,e$)
have broad upflows and narrow downflows as a
consequence of the compressible motions.  Near the equator the
convection is aligned largely in the north-south direction, and
these broad fronts sweep through the domain in a prograde fashion.
The strongest downflows penetrate to the bottom of the convection zone;
the weaker flows are partially truncated by the strong zonal flows
of differential rotation.  In the polar regions the convection is more
isotropic and cyclonic.  There the networks of downflow lanes surround upflows
and both propagate in a retrograde fashion.

The convection establishes a prominent differential rotation profile by
redistributing angular momentum and entropy, building gradients
in latitude of angular velocity and temperature.  
Figures~\ref{fig:case_D3_patterns}$b,f$ show the mean angular velocity
$\Omega(r,\theta)$ for cases D3 and H3, revealing a solar-like
structure with a prograde (fast) equator and retrograde (slow) pole.
Figures~\ref{fig:case_D3_patterns}$c,g$ present in turn radial
cuts of $\Omega$ at selected latitudes, which are useful as we consider
the angular velocity patterns realized here with faster rotation. 
These $\Omega(r,\theta)$ profiles are averaged in azimuth (longitude)
and time over a period of roughly 200~days. 
Contours of constant angular velocity are aligned nearly
on cylinders, influenced by the Taylor-Proudman theorem.  

In the Sun, helioseismology has revealed that the contours of angular velocity
are aligned almost on radial lines rather than on cylinders.  The tilt
of $\Omega$ contours in the Sun may be due in part to the thermal
structure of the solar tachocline, as first found in the mean-field
models of \cite{Rempel_2005} and then in 3D simulations  of global-scale
convection by \cite{Miesch_et_al_2006}.  In those computations, it was
realized that introducing a weak latitudinal gradient of entropy at the
base of the convection zone, consistent with a thermal wind balance in
a tachocline of shear, can serve to tilt the $\Omega$ contours
toward a more radial alignment without significantly changing either
the overall $\Omega$ contrast with latitude or the convective
patterns.  \cite{Ballot_et_al_2007} explored the consequences of such
a boundary condition in one of their simulations of young, rapidly
rotating suns with deep convection zones and found that the
effects on the differential rotation were similar to those found in
\cite{Miesch_et_al_2006}.   We expect similar behavior here, but at present
observations of rapidly rotating stars only measure differential
rotation at the surface and do not offer constraints on either the
existence of tachoclines in young suns or the nature of their internal
differential rotation profiles.  As such, we have
neglected the possible tachoclines of penetration and shear entirely
in these models and instead adopt the simplification of imposing a
constant radial entropy gradient at the bottom of the convection
zone.   

\begin{figure*}
  \includegraphics[width=\linewidth]{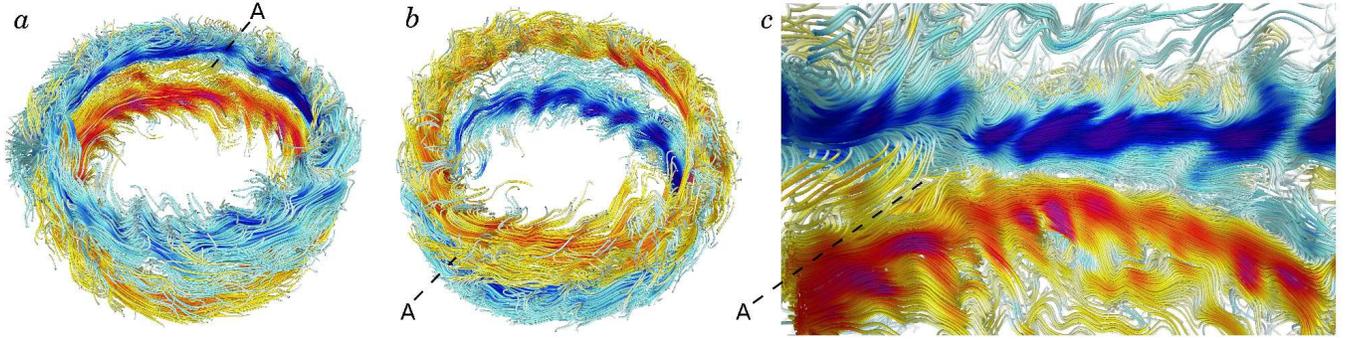}
  \caption{Field line tracings of magnetic wreaths in case~D3.  
    $(a)$~Snapshot of two wreaths in full volume at same instant
    as in Fig.~\ref{fig:case_D3_many_depths}.  Lines
    trace the magnetic fields, color denoting the amplitude and
    polarity of the longitudinal field $B_\phi$ (red, positive;
    blue, negative).  Magnetic field threads in and out of the wreaths,
    connecting the two opposite polarity structures across the equator
    (i.e., region A) and to the polar  regions where the magnetic field
    is wound up by the cyclonic convection.
    $(b)$~Same snapshot showing south polar region.  
    $(c)$~Zoom in on region A showing the complex interconnections
    across the equator between the two wreaths and to high
    latitudes.  Convective flows create the distinctive waviness
    visible in all three images.
  \label{fig:case_D3_field_lines}}
\end{figure*}

\begin{deluxetable}{lcccc}
   \tabletypesize{\footnotesize}
    \tablecolumns{5}
    \tablewidth{0pt}  
    \tablecaption{Near-surface $\Delta \Omega$
    \label{table:delta_omega}}
  \tablehead{\colhead{Case}  &  
      \colhead{$\Delta \Omega_\mathrm{lat}$} &
      \colhead{$\Delta \Omega_\mathrm{r}$} &
      \colhead{$\Delta \Omega_\mathrm{lat}/\Omega_\mathrm{eq}$} &
      \colhead{Epoch}
   }
   \startdata
    D3                       & 1.18 & 0.71 & 0.137 & 2010-6980 \\
    H3                       & 2.22 & 0.94 & 0.246 &  - \\
    \enddata
 \tablecomments{Angular velocity shear in units of $\mu
   \mathrm{rad}\: s^{-1}$, with $\Delta \Omega_\mathrm{lat}$ measured near the surface
   ($0.97R_\odot$) and $\Delta \Omega_\mathrm{r}$ measured across the
   full shell at the equator.  The relative latitudinal shear $\Delta
   \Omega_\mathrm{lat}/\Omega_\mathrm{eq}$ is also measured at the
   same point near the surface.  For the dynamo case, these
   measurements are taken over the indicated range of days.  
   Case~D3 shows slow variations in $\Delta \Omega_\mathrm{lat}$ over
   periods of about 2000~days.
   The hydrodynamic case is averaged for roughly 300 days and shows no
   systematic variation on longer timescales.}
\end{deluxetable}

The differential rotation achieved is stronger in our hydrodynamic
case~H3 than in our dynamo case~D3.
This can be quantified by measurements of the 
latitudinal angular velocity shear $\Delta \Omega_\mathrm{lat}$.  Here, as in
\cite{Brown_et_al_2008}, we define $\Delta \Omega_\mathrm{lat}$ as the shear near
the surface between the equator and a high latitude, say $\pm 60^\circ$
\begin{equation}
  \Delta \Omega_\mathrm{lat} = \Omega_\mathrm{eq} - \Omega_{60},
  \label{eq:absolute_contrast}
\end{equation}
and the radial shear $\Delta \Omega_\mathrm{r}$ as the angular
velocity shear between the surface and bottom of the convection zone
near the equator 
\begin{equation}
  \Delta \Omega_\mathrm{r} = \Omega_{0.97R_\odot} - \Omega_{0.72R_\odot}.
  \label{eq:absolute_radial_contrast}
\end{equation}
We further define the relative shear as 
$\Delta \Omega_\mathrm{lat}/\Omega_\mathrm{eq}$.
In both definitions, we average the measurements of $\Delta \Omega$
in the northern and southern hemispheres, as the rotation profile is
often slightly asymmetric about the equator.
Case~H3 achieves an absolute contrast $\Delta \Omega_\mathrm{lat}$ of
2.22~$\mu~\mathrm{rad}\:\mathrm{s}^{-1}$ (352~nHz) and a relative contrast of 0.247.
The strong global-scale magnetic fields realized in the dynamo case~D3
serve to diminish the differential rotation.  As such, this case
achieves an absolute contrast $\Delta \Omega_\mathrm{lat}$ of only 
1.18~$\mu~\mathrm{rad}\:\mathrm{s}^{-1}$ (188~nHz) and a relative contrast of 0.137.
This results from both a slowing of the equatorial rotation rate and
an increase in the rotation rate in the polar regions.  These results
are quoted in Table~\ref{table:delta_omega}.

The meridional circulations realized in the dynamo case~D3 are very
similar to those found in its hydrodynamic progenitor (case~H3).  As
illustrated in Figures~1$d, h$, the circulations are multi-celled in
radius and latitude.  The cells are strongly aligned with the rotation
axis, though some flows along the inner and outer boundaries cross the
tangent cylinder and serve to weakly couple the polar regions to the
equatorial convection.  Flows of meridional circulation are slightly
stronger in the dynamo cases than in the purely hydrodynamic cases,
though both cases have weaker flows than are found in simulations
rotating at the solar rate.  Thus, as found in \cite{Brown_et_al_2008}, the
flows of meridional circulation appear to weaken with more rapid
rotation.  The multi-celled nature of these meridional circulations
may hold implications for flux transport dynamo models
\citep[e.g,][]{Jouve&Brun_2007}.  Recent mean-field dynamo models are also
beginning to explore the implications of weaker and multi-celled
meridional circulations for dynamo action in more rapidly rotating suns
\citep[e.g.,][]{Jouve_et_al_2009}.

\subsection{Kinetic and Magnetic Energies}
Convection in these rapidly rotating dynamos is responsible for
building the differential rotation and the magnetic fields.  In a
volume averaged sense, the energy contained in the magnetic fields
in case~D3 is about 10\% of the kinetic energy.  About 35\%
of this kinetic energy is contained in the fluctuating convection
(CKE) and about 65\% in the differential rotation (DRKE), whereas
the weaker meridional circulations contain only a small portion
(MCKE).  The magnetic energy is split between the contributions from
fluctuating fields (FME), involving roughly 53\% of the total magnetic
energy, and the energy of the mean toroidal fields (TME) that are
43\% of the total.  The energy contained in the mean poloidal fields
(PME) is only 4\% of the total magnetic energy.  These energies are defined as
\begin{eqnarray}
  \mathrm{CKE}  &=& \frac{1}{2}\bar{\rho}\Big[\left(v_r - \langle v_r \rangle\right)^2+\left(v_\theta - \langle v_\theta \rangle\right)^2+\nonumber\\
                & & \qquad \left(v_\phi - \langle v_\phi \rangle\right)^2\Big], \\
  \mathrm{DRKE} &=& \frac{1}{2}\bar{\rho}\langle v_\phi \rangle^2, \\
  \mathrm{MCKE} &=& \frac{1}{2}\bar{\rho}\Big(\langle v_r \rangle^2 + \langle v_\theta \rangle^2 \Big),\\
\end{eqnarray}
\begin{eqnarray}
  \mathrm{FME}  &=& \frac{1}{8\pi}\Big[\left(B_r - \langle B_r \rangle\right)^2+\left(B_\theta - \langle B_\theta \rangle\right)^2+\nonumber\\
                & & \qquad \left(B_\phi - \langle B_\phi \rangle\right)^2\Big], \\
  \mathrm{TME}  &=& \frac{1}{8\pi}\langle B_\phi \rangle^2, \\
  \mathrm{PME}  &=& \frac{1}{8\pi}\Big(\langle B_r \rangle^2 + \langle B_\theta \rangle^2 \Big).
\end{eqnarray}
where angle brackets denote an average in longitude.

These results are in contrast to our previous simulations of the solar
dynamo, where the mean fields contained only about 2\% of the magnetic
energy and the fluctuating fields contained nearly 98\%
\citep{Brun_et_al_2004}.  In simulations of the solar dynamo that
included a stable tachocline at the base of the convection zone
\citep{Browning_et_al_2006}, the energy of the mean fields in the
tachocline can exceed the energy of the fluctuating fields there 
by about a factor of three, though 
the fluctuating fields still dominate the magnetic
energy budget within the convection zone itself.  Simulations of dynamo
activity in the convecting cores of A-type stars
\citep{Brun_et_al_2005} achieved similar results.  There in the
stable radiative zone the energies of the mean fields were able to
exceed the energy contained in the fluctuating fields, but in the
convecting core the fluctuating fields contained roughly 95\% of the
magnetic energy.  Simulations of dynamo action in fully-convective
M-stars do however show high levels of magnetic energy in the mean
fields \citep{Browning_2008}. In those simulations the fluctuating
fields still contain much of the magnetic energy,
but the mean toroidal fields possess about 18\% of the total
throughout most of the stellar interior. 
In our rapidly rotating suns, the mean fields
comprise a significant portion of the magnetic energy in the
convection zone and are as important as the fluctuating fields.

Convection is similarly strong in both rapidly rotating cases, and
CKE is similar in magnitude.  The differential rotation in the dynamo
case is much weaker than in the hydrodynamic progenitor,
and DRKE has decreased by about a factor of five.  Meridional
circulations are comparably weak in both cases.

\begin{deluxetable}{lccccccc}
   \tabletypesize{\footnotesize}
    \tablecolumns{7}
    \tablewidth{0pt}  
    \tablecaption{Energies
    \label{table:energies}}
    \tablehead{\colhead{Case}  &  
      \colhead{CKE} &
      \colhead{DRKE} &
      \colhead{MCKE} &
      \colhead{FME} &
      \colhead{TME} &
      \colhead{PME} 
   }
   \startdata
    D3                       & $2.31$ & $4.35$ & $0.010$ &
                               $0.36$ & $0.29$ & $0.029$ \\ 

    H3                       & $2.56$ & $22.2\phn\phn$ & $0.012$ &
                                -     &   -    & - \\ 

    \enddata
 \tablecomments{Volume-averaged energy densities relative to the
    rotating coordinate system.  Kinetic energies are shown for
    convection (CKE), differential rotation (DRKE) and meridional
    circulations (MCKE).  Magnetic energies are shown for fluctuating
    magnetic fields (FME), mean toroidal fields (TME) and mean
    poloidal fields (PME).  All energy densities are reported in units of
    $10^{6} \mathrm{erg}\:\mathrm{cm}^{-3}$ and are 
    averaged over 1000 day periods.}
\end{deluxetable}

\begin{figure*}
  \includegraphics{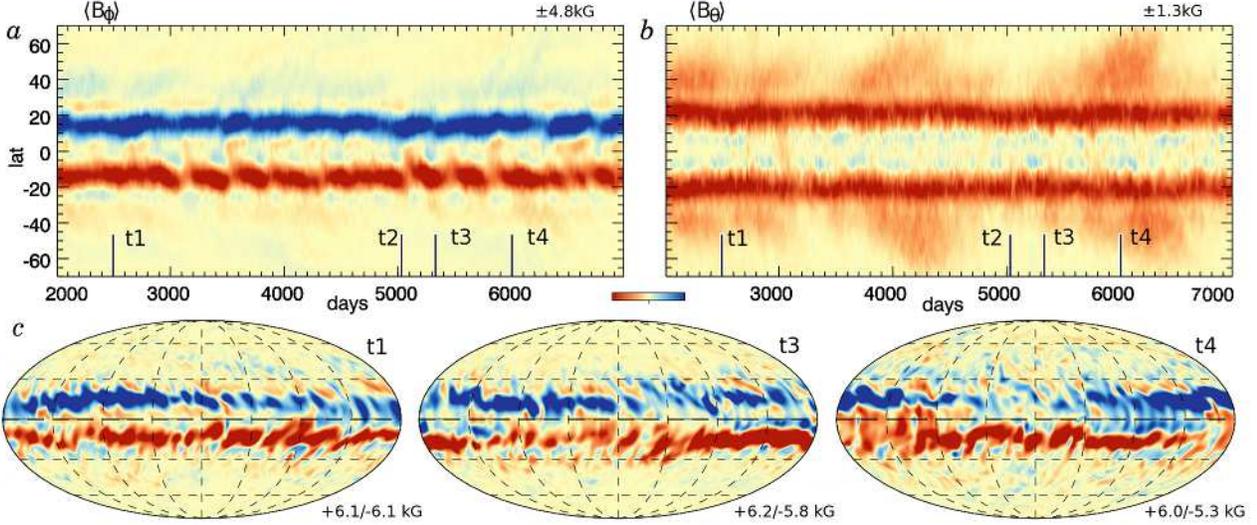}
  \caption{Persistent wreaths of magnetism in case~D3.  $(a)$~Time-latitude plots of
  azimuthally-averaged longitudinal field $\langle B_\phi \rangle$ 
  at mid-convection zone ($0.85R_\sol$) in a view spanning latitudes from
  $\pm 70^\circ$, with scaling values
  indicated.  The two wreaths of opposite polarity persist for more
  than 4000 days. $(b)$~Mean colatitudinal magnetic field 
  $\langle B_\theta \rangle$ at
  mid-convection zone over same interval.  $(c)$~Snapshots of $B_\phi$
  in Mollweide projection at mid-convection zone, shown for three
  times indicated in $a,b$.  The wreaths maintain constant polarity
  over long time intervals, but still show variation as they interact
  with the convection.  Time t2 corresponds to the snapshot in 
  Fig.~\ref{fig:case_D3_many_depths}$b$.
  \label{fig:case_D3_time_evolution}}
\end{figure*}

\section{Wreaths of Magnetism}
\label{sec:wreaths}

These rapidly rotating dynamos produce striking magnetic structures in the midst of
their turbulent convection zones.  The magnetic field is organized
into large banded, wreath-like structures positioned near the equator
and spanning the depth of the convection zone.  These wreaths are
shown for case~D3 at two depths in the convection zone in
Figure~\ref{fig:case_D3_many_depths}.    
The dominant component of the magnetic wreaths is the strong longitudinal
field $B_\phi$, with each wreath possessing its own polarity.  The average
strength of the longitudinal field at mid-convection zone is $\pm 7$~kG 
and peak field strengths there reach roughly $\pm 26$~kG.  
Threaded throughout the wreaths are weaker radial and latitudinal magnetic
fields, which connect the two structures across the equator and also
to the high-latitude regions.  

These wreaths of magnetism survive despite being embedded in vigorous
convective upflows and downflows.  The
convective flows leave their imprint on the magnetic structures,  
with individual downflow lanes entraining the
magnetic field, advecting it away, and stretching it into $B_r$
while leaving regions of locally reduced $B_\phi$. The slower upflows
carry stronger $B_\phi$ up from the depths.  Where the magnetism is
particularly strong the convective flows are disrupted.
Meanwhile, where the convective flows are strongest, the longitudinal
magnetic field is weakened and appears to 
vanish.  In  reality, the magnetic wreaths here are
diving deeper below the mid-convection zone, apparently pumped
down by the pummeling action of the strong downflows.  

The deep structure of these wreaths is revealed by field line
tracings throughout the volume, shown in
Figure~\ref{fig:case_D3_field_lines} for the same instant in time.
The wreaths are topologically leaky structures, with magnetic field
lines threading in and out of the surrounding convection.  The
wreaths are connected to the high-latitude (polar) convection, and on
the poleward edges they show substantial winding from the highly
vortical convection found there.  This occurs in both the northern and
southern hemispheres, as shown in two views at the same instant
(north, Fig.~\ref{fig:case_D3_field_lines}$a$ and south,
Fig.~\ref{fig:case_D3_field_lines}$b$).  It is here that the
global-scale poloidal field is being regenerated by the coupling of
fluctuating velocities and fluctuating fields. Magnetic fields cross the
equator, tying the two wreaths together at many locations
(Fig.~\ref{fig:case_D3_field_lines}$c$).  The strongest convective
downflows leave their imprint on the wreaths as regions where the
field lines are dragged down deeper into the convection zone, yielding
a wavy appearance to the wreaths as a whole.

\begin{figure}
  \vspace{1cm}
  \includegraphics[width=9cm]{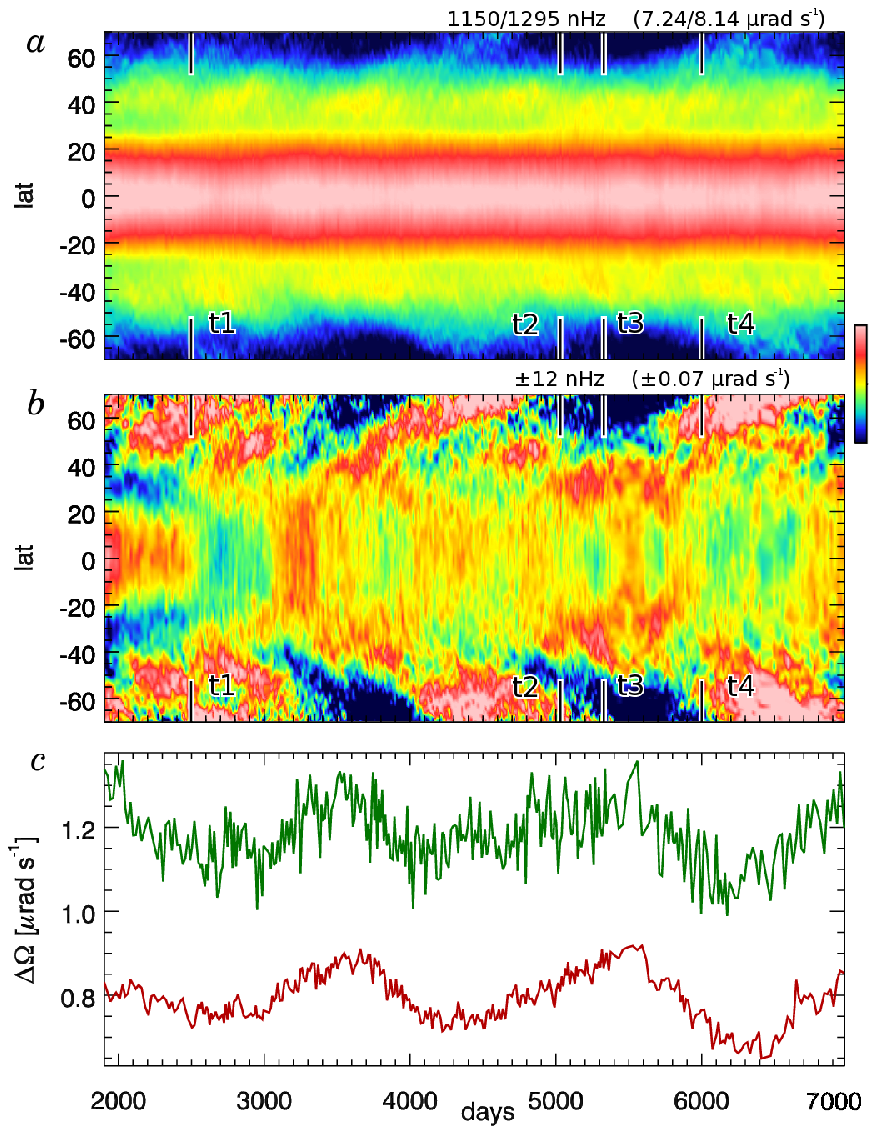}
  \caption{Differential rotation in case~D3.  $(a)$~Angular velocity
  $\Omega$ at mid-convection zone ($0.85R_\sol$), with ranges in both nHz
  and $\mu\mathrm{rad}\:\mathrm{s}^{-1}$.  The equator is
  fast while the poles rotate more slowly.  
  $(b)$~Temporal variations are emphasized by subtracting the
  time-averaged profile of $\Omega(r,\theta)$, revealing
  speedup structures at high latitudes and pulses of fast and slow
  motion near the equator.  $(c)$~Angular velocity
  shear $\Delta \Omega_\mathrm{lat}$ (eq.~\ref{eq:absolute_contrast}) near the
  surface (upper~curve, green) and at mid-convection zone (lower, red).
  \label{fig:case_D3_DR}
  }
\end{figure}

\subsection{Wreaths Persist for Long Epochs}
The wreaths of magnetism built in case~D3 persist for
long periods of time, with little change in strength and no reversals
in global-scale polarity for as long as we have pursued
these calculations.  The long-term stability of the wreaths realized
by the dynamo of case~D3 is shown in Figure~\ref{fig:case_D3_time_evolution}.
Here the azimuthally-averaged longitudinal field $\langle B_\phi \rangle$
and colatitudinal field $\langle B_\theta \rangle$ are shown at
mid-convection zone at a point after the dynamo has equilibrated and
for a period of roughly 5000~days (i.e.,~several ohmic diffusion times).  
During this interval there is little
change in either the amplitude or structure of the mean fields.  This
is despite the short overturn times of the convection (10-30
days) or the rotation period of the star ($\sim 9$~days).  The ohmic
diffusion time at mid-convection zone is approximately 1300 days.

Though the mean (global-scale) fields are roughly steady in nature
(Figs.~\ref{fig:case_D3_time_evolution}$a,b$), the magnetic
field interacts strongly with the convection on smaller scales. 
Several samples of longitudinal field $B_\phi$ are shown in full Mollweide projection at
mid-convection zone (Fig.~\ref{fig:case_D3_time_evolution}$c$).  
The magnetic fields are clearly reacting on short time scales to the
convection but the wreaths maintain their coherence.  

There are also some small but repeated variations in the global-scale magnetic fields.
Visible in Figure~\ref{fig:case_D3_time_evolution}$b$ are
events where propagating structures of $\langle B_\theta \rangle$ reach toward
higher latitudes over periods of about 1000~days (i.e., from day 3700
to day 4500 and from day 5600 to day 6400).  These are accompanied by
slight variations in the volume-averaged magnetic energy densities and
the comparable kinetic energy of the differential rotation.
These variations are also visible in the differential rotation itself,
as shown in Figure~\ref{fig:case_D3_DR}.  The differential rotation
is fairly stable, though some time variation is visible at high
latitudes.  This is better revealed (Fig.~\ref{fig:case_D3_DR}$b$)
by subtracting the time-averaged profile of $\Omega$ at each
latitude, revealing the temporal variations about this mean.   
In the polar regions above $\pm 40^\circ$ latitude, speedup features
move poleward over 500 day periods.  These features
track similar structures visible in the mean magnetic fields
(Fig.~\ref{fig:case_D3_time_evolution}$b$).   
The bands of velocity speedup bear some resemblance to the poleward
branch of torsional oscillations observed in the solar convection
zone over the course of a solar magnetic activity cycle
\citep[e.g.,][]{Thompson_et_al_2003, Howe_2009}, though here they
propagate to higher latitudes on a  shorter time scale.

The temporal variations of the angular velocity contrast in latitude
$\Delta \Omega_\mathrm{lat}$ are shown for this period in
Figure~\ref{fig:case_D3_DR}$c$.  At mid-convection zone (sampled by
red line) the variations in $\Delta \Omega_\mathrm{lat}$ are
modest, varying by roughly 8\%. 
Near the surface (green line) $\Delta \Omega_\mathrm{lat}$ shows
similar variations with amplitudes of about 6\%.  The near-surface
values of $\Delta \Omega_\mathrm{lat}$ are reported in
Table~\ref{table:delta_omega}, averaged over this entire period.

These evolving structures of magnetism and faster and slower
differential rotation appear to be the
first indications of behavior where the mean fields themselves begin to
wax and wane substantially in strength.  As the magnetic Reynolds
number is increased, by either decreasing the magnetic diffusivity
$\eta$ or by increasing the rotation rate of the star $\Omega_0$, 
this time varying behavior becomes more prominent
and can even result in organized changes in the global-scale polarity.
Such behavior is evident in a number of our dynamo simulations and
will be reported on in a subsequent paper.

\section{Creating Magnetic Wreaths}
\label{sec:dynamo_production}
The magnetic wreaths formed in case~D3 are dominated by strong 
mean longitudinal field components and show little variation in time. 
To understand the physical processes responsible
for maintaining these magnetic wreaths, we examine the terms arising in
the time- and azimuth-averaged induction equation for case~D3.

\begin{figure*}
  \includegraphics{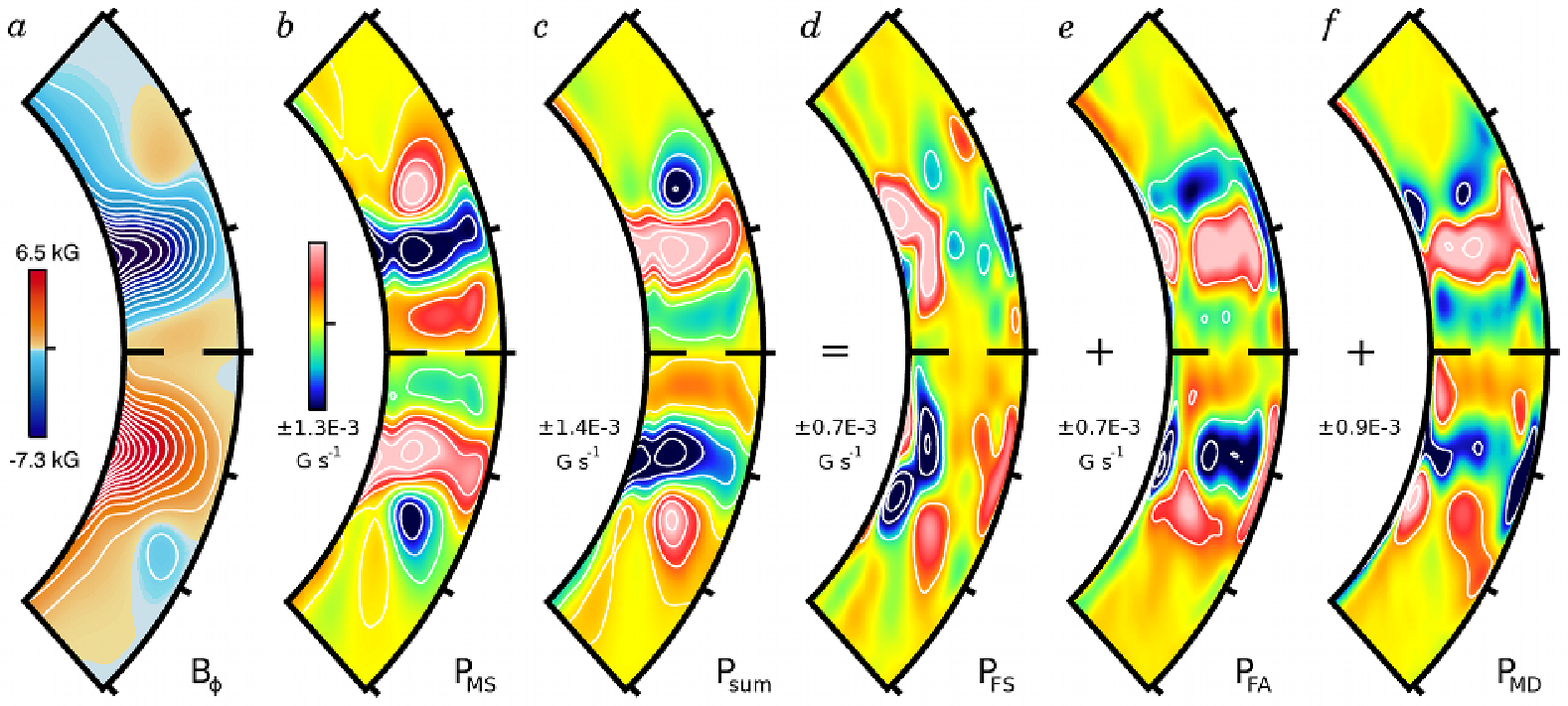}
  \caption{Generation of mean toroidal magnetic field in case~D3.  The
    view is from $\pm 45^\circ$ latitude to emphasize the
    equatorial regions.  $(a)$~Mean toroidal field $\langle B_\phi
    \rangle$ with wreaths strongly evident.  $(b)$~Production by
    $P_\mathrm{MS}$ serves to build $\langle B_\phi \rangle$.
    This rate term generally matches the sense of $\langle B_\phi \rangle$,
    thus being negative (blue in colorbar, with ranges indicated) in the core of the
    northern wreath and positive (red) in that of the southern wreath.
    $(c)$~Destruction of mean toroidal field is achieved by the sum of
    the two fluctuating (turbulent) induction terms and the ohmic
    diffusion $\left(P_\mathrm{FS} + P_\mathrm{FA} + P_\mathrm{MD}\right)$.
    This sum clearly has opposite sense and similar magnitude to $P_\mathrm{MS}$.  
    We break out these three destruction terms in the following panels.
    $(d)$~Fluctuating (turbulent) shear
    $P_\mathrm{FS}$ is strongest near the high-latitude side of each
    wreath, and $(e)$~fluctuating (turbulent) advection $P_\mathrm{FA}$ is strongest
    in the cores of the wreaths.  The sum of these terms
    $\left(P_\mathrm{FS} + P_\mathrm{FA}\right)$ is
    responsible for about half the destructive balance, with the
    remainder coming from $(f)$~the mean ohmic diffusion $P_\mathrm{MD}$.
    Some differences arise in the boundary layers at top and bottom.
    \label{fig:D3_bphi_production}}
\end{figure*}

\subsection{Maintaining Wreaths of Toroidal Field}
We begin our analysis by exploring the maintenance of the mean
toroidal field $\langle B_\phi \rangle$.  Here it is helpful to break
the induction term from equation~(\ref{eq:induction}) into
contributions from shear, advection and compression, namely
\begin{eqnarray}
  \vec{\del} &\cross &\left(\vec{v} \cross \vec{B} \right) = \nonumber \\
  & &\underbrace{\left(\vec{B} \cdot \vec{\del} \right) \vec{v}}_{\mbox{shear}}
  - \underbrace{\left(\vec{v} \cdot \vec{\del} \right)  \vec{B}}_{\mbox{advection}}
  - \underbrace{\vec{B} \left(\vec{\del} \cdot \vec{v} \right)}_{\mbox{compression}}.
\end{eqnarray}
Details of this decomposition are given in the~Appendix.

The evolution of the mean longitudinal (toroidal) field $\langle B_\phi \rangle$
is described symbolically in equation~(\ref{eq:mean induction}), with
individual terms defined in equation~(\ref{eq:mean terms}).
When we analyze these terms in case~D3, we find that  $\langle
B_\phi \rangle$ is produced by the shear of differential rotation and
is dissipated by a combination of turbulent induction and ohmic
diffusion.  This balance can be restated as  
\begin{equation}
  \frac{\partial \langle B_\phi \rangle}{\partial t} \approx 
  P_\mathrm{MS} + \left(P_\mathrm{FS} + P_\mathrm{FA} + P_\mathrm{MD}\right)
  \approx 0\thinspace ,
\end{equation}
with $P_\mathrm{MS}$ representing production by the mean shearing flow of
differential rotation, 
$P_\mathrm{FS}$ by fluctuating shear, $P_\mathrm{FA}$ by fluctuating
advection, and $P_\mathrm{MD}$ by mean ohmic diffusion.
Those terms are in turn 
\begin{eqnarray}
  \label{eq:P_DR}
  P_\mathrm{MS} &=& \phn\left(\langle \vec{B} \rangle \cdot \vec{\del} \right) 
    \langle \vec{v} \rangle \big|_\phi, \\
  \label{eq:P_turb}
  P_\mathrm{FS} &=& \phn\left\langle \left(\vec{B}' \cdot \vec{\del} \right)  
    \vec{v'} \right\rangle\big|_\phi, \\
  P_\mathrm{FA} &=& - \left\langle \left(\vec{v}' \cdot \vec{\del} \right) 
    \vec{B'} \right\rangle\big|_\phi, \\
  \label{eq:P_diff}
  P_\mathrm{MD} &=& -\vec{\del} \cross \eta \vec{\del} \cross \langle
    \vec{B} \rangle \big|_\phi,
\end{eqnarray}
where brackets again indicate an azimuthal average and primes
indicate fluctuating terms: $\vec{v}' = \vec{v} - \langle \vec{v} \rangle$.
The detailed implementation of these terms  is presented for our
spherical geometry in equations~(\ref{eq:MS phi}-\ref{eq:MD phi}).
These terms are illustrated in Figure~\ref{fig:D3_bphi_production}
for case~D3, averaged over a 450~day interval from day~6450~to~6900.  

The structure of $\langle B_\phi \rangle$ is shown in
Figure~\ref{fig:D3_bphi_production}$a$.  The shearing flows of
differential rotation $P_\mathrm{MS}$
(Fig.~\ref{fig:D3_bphi_production}$b$) act almost everywhere to
reinforce the mean toroidal field.  Thus the polarity of this production term
generally matches that of $\langle B_\phi \rangle$.  This production is balanced by
destruction of mean field arising from both turbulent induction and ohmic diffusion
(sum shown in Fig.~\ref{fig:D3_bphi_production}$c$).  The individual
profiles of $P_\mathrm{FS}$, $P_\mathrm{FA}$ and $P_\mathrm{MD}$ are
presented in turn in Figures~\ref{fig:D3_bphi_production}$d,e,f$.
The terms from turbulent induction ($P_\mathrm{FS}$ and
$P_\mathrm{FA}$) contribute to roughly half of the total balance, with
the remainder carried by ohmic diffusion of the mean fields
($P_\mathrm{MD}$).  In the core of the wreaths, removal of mean
toroidal field is largely accomplished
by fluctuating advection $P_\mathrm{FA}$
(Fig.~\ref{fig:D3_bphi_production}$e$) and mean ohmic diffusion
$P_\mathrm{MD}$ (Fig.~\ref{fig:D3_bphi_production}$f$), with the latter also important near the upper boundary.
Turbulent shear becomes strongest
near the bottom of the convection zone and in the
regions near the high-latitude side of each wreath.
Thus $P_\mathrm{FS}$ (Fig.~\ref{fig:D3_bphi_production}$d$) becomes the dominant member of the triad of terms
seeking to diminish the mean toroidal field there. 
We find that the mean poloidal field is regenerated in roughly the
same region.

In the analysis presented in Figure~\ref{fig:D3_bphi_production} we have
neglected the advection of $\langle B_\phi \rangle$ 
by the meridional circulations (shown in the~Appendix as
$P_\mathrm{MA}$), which we find plays a very small role
in the overall balance.  We have also neglected the amplification of
$\langle B_\phi \rangle$ by compressibility effects (the~Appendix,
$P_\mathrm{MC}$ and $P_\mathrm{FC}$), though it does
contribute slightly to reinforcing the underlying mean fields within
the wreaths.

To summarize, the mean toroidal fields are built through an
$\Omega$-effect, where production by the mean shearing flow of
differential rotation ($P_\mathrm{MS}$) builds the underlying
$\langle B_\phi \rangle$.  In the statistically steady state achieved, this
production is balanced by a combination of turbulent
induction ($P_\mathrm{FS} + P_\mathrm{FA}$) and ohmic diffusion of the
mean fields ($P_\mathrm{MD}$).

\subsection{Maintaining the Poloidal Field}
The production of mean poloidal field is achieved through a slightly
different balance, with turbulent induction producing poloidal field and
ohmic diffusion acting to dissipate it.  The mean flows play
little role in the overall balance.  This balance is clarified if we
represent the mean poloidal field by its vector potential 
$\langle A_\phi \rangle$, where  
\begin{equation}
    \langle \vec{B}_\mathrm{pol} \rangle = \langle B_r \rangle \vec{\hat{r}} +
    \langle B_\theta \rangle \vec{\hat{\theta}} 
    = \vec{\del} \cross \left\langle A_\phi \vec{\hat{\phi}} \right\rangle,
\end{equation}
as discussed in the~Appendix.
We recast the induction equation (\ref{eq:induction}) in terms of
the poloidal vector potential by uncurling the equation once, obtaining
\begin{equation}
    \frac{\partial\langle\vec{A_\phi}\rangle}{\partial t} = 
        \left\langle \vec{v} \cross \vec{B}\right\rangle{\big|}_\phi 
	- \eta \vec{\del} \cross \left \langle \vec{B} \right\rangle{\big|}_\phi,
\end{equation}
which is also equation~(\ref{eq:A induction}) in the~Appendix.
The first term is the electromotive force (emf) arising from the
coupling of flows and magnetic fields, and the second term is
the ohmic diffusion.  These can be
decomposed into contributions from mean and fluctuating
components, as shown symbolically in equation~(\ref{eq:A induction symbolic}).

In case~D3 we find that the mean poloidal vector potential 
$\langle A_\phi \rangle$ is produced by the fluctuating (turbulent) emf and is
dissipated by ohmic diffusion
\begin{equation}
  \label{eq:poloidal balance}
  \frac{\partial \langle A_\phi \rangle}{\partial t} \approx 
  E_\mathrm{FI} +
  E_\mathrm{MD} \approx 0\thinspace.
\end{equation}
with $E_\mathrm{FI}$ the emf arising from fluctuating flows and
fluctuating fields, and contributing to the mean induction.  The
$E_\mathrm{MD}$ is the emf arising from mean
ohmic diffusion.  These terms are
\begin{eqnarray}
  \label{eq:E_FI}
  E_\mathrm{FI} &=& \langle \vec{v'}\cross\vec{B'} \rangle{\big|}_\phi
  = \langle v_r' B_\theta' \rangle - 
    \langle v_\theta' B_r' \rangle, \\
  E_\mathrm{MD} &=& -\eta \vec{\del} \cross \langle \vec{B} \rangle{\big|}_\phi.
\end{eqnarray}
The contribution arising from the omitted term $E_\mathrm{MI}$ 
(see~eq.~\ref{eq:A MI phi}), related to the emf of mean flows and mean
fields, is smaller than these first two by more than an order
of magnitude.  Additionally, $E_\mathrm{MI}$ has a complicated spatial
structure which does not appear to act in a coherent fashion within
the wreaths to either build or destroy mean poloidal field.

\begin{figure}
  \includegraphics[width=9cm]{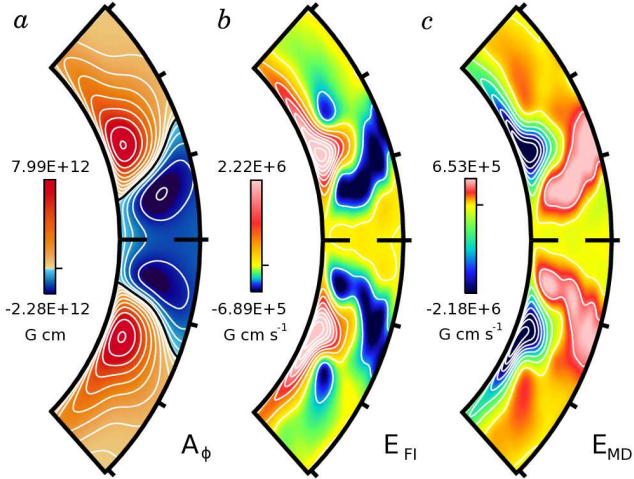}
  \caption{Production of mean poloidal vector potential $\langle
    A_\phi \rangle$ in case~D3, with view restricted to $\pm 45^\circ$
    latitude.  $(a)$~Mean poloidal vector potential 
    $\langle A_\phi \rangle$, with sense denoted by color (red,
    clockwise; blue, counter-clockwise).
    $(b)$~The fluctuating (turbulent) emf
    $E_\mathrm{FI}$ acts to build the vector potential.  
    This term is strongest near the bottom of the convection zone and
    the poleward side of the wreaths.  
    $(c)$~Mean ohmic diffusion $E_\mathrm{MD}$ acts everywhere in
    opposition to $E_\mathrm{FI}$.  
    The cores of the wreaths are positioned at roughly
    $\pm15^\circ$ latitude (Fig.~\ref{fig:D3_bphi_production}$a$).
  \label{fig:D3 poloidal production}}
  \vskip0.1truein
\end{figure}

The mean vector potential $\langle A_\phi \rangle$ is shown in Figure~\ref{fig:D3 poloidal
production}$a$, with poloidal field lines represented by the overlying contours.  
The mean radial magnetic field $\langle B_r \rangle$ is
about $\pm 1$~kG in the cores of the wreaths, whereas the mean colatitudinal
field $\langle B_\theta \rangle$ has an amplitude of roughly $-2$~kG
(thus directed northward in both hemispheres),
concentrated near the bottom of the convection zone.

The production of $\langle A_\phi \rangle$ by the fluctuating (turbulent) emf
$E_\mathrm{FI}$ is shown in Figure~\ref{fig:D3 poloidal production}$b$.  
Here too we average over the same 450~day interval.
This term generally acts to reinforce the
existing poloidal field, having the same sense as the underlying
vector potential in most regions.  It is strongest near the bottom of
the convection zone and is concentrated at the poleward  side of
each wreath.  This is similar, though not identical, to the structure
of destruction of mean toroidal field by fluctuating shear $P_\mathrm{FS}$
(Fig.~\ref{fig:D3_bphi_production}$d$).  It suggests that mean 
toroidal field is here being converted into mean poloidal field by the
fluctuating flows.

There are two terms that contribute to $E_\mathrm{FI}$, as shown in equation~(\ref{eq:E_FI}).
Much of that fluctuating emf arises from
correlations between fluctuating latitudinal flows and radial fields
$\langle -v_\theta'B_r' \rangle$, which follows the structure of $E_\mathrm{FI}$
(Fig.~\ref{fig:D3 poloidal production}$b$) closely.  The contribution from
fluctuating radial flows and colatitudinal fields $\langle
v_r'B_\theta'\rangle$ is more complex in structure.  
Near $\pm 20^\circ$ latitude, this term reinforces
$\langle -v_\theta'B_r' \rangle$, but acts against it at
higher latitudes and thus diminishes the overall amplitude of
$E_\mathrm{FI}$. 
The mean ohmic diffusion $E_\mathrm{MD}$ (Fig.~\ref{fig:D3
  poloidal production}$c$), almost entirely balances the production of
$\langle A_\phi \rangle$ by $E_\mathrm{FI}$.

This shows that our mean poloidal magnetic field is maintained by the
fluctuating (turbulent) emf and is destroyed by ohmic diffusion.  In
mean-field dynamo theory, this is often parametrized by an
``$\alpha$-effect.''  Now we turn to interpretations within that framework.

\begin{figure}
  \includegraphics[width=9cm]{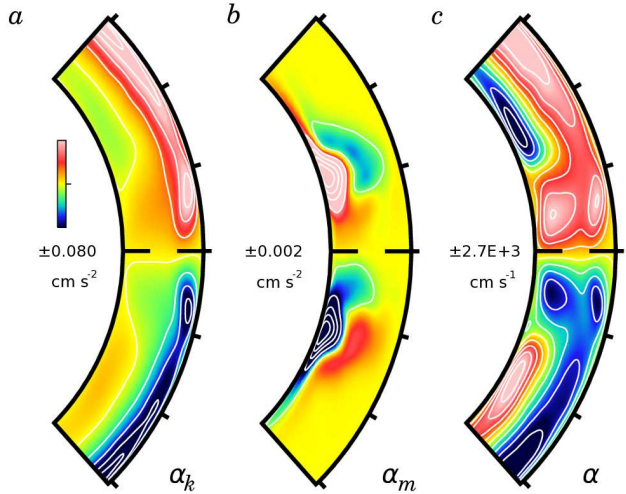}
  \caption{Estimating the mean-field $\alpha$-effect from case~D3.  
    Shown are the $(a)$~kinetic and $(b)$~magnetic contributions to
    the $\alpha$-effect as defined in eqs.~(\ref{eq:mean-field
      alpha}-\ref{eq:mean-field alpha_m}). 
    $(c)$~Mean-field $\alpha$, constructed by combining
    $\alpha_k$~and~$\alpha_m$ with a turbulent correlation time $\tau$.
  \label{fig:mean_field_alpha}}
  \vspace{0.2cm}
\end{figure}

\section{Exploring Mean-Field Interpretations}
\label{sec: mean-field}

Many mean-field theories assert that the production of mean poloidal
field is likely to arise from the fluctuating emf.  This process is often
approximated with an $\alpha$-effect, where it is proposed that the
sense and amplitude of the emf scales with the mean toroidal field
\begin{equation}
  \label{eq:mean field emf}
    \langle \vec{v'}\cross\vec{B'} \rangle = \alpha \langle \vec{B} \rangle,
\end{equation}
where $\alpha$ can be either a simple scalar or may be related to the
kinetic and magnetic (current) helicities.  In isotropic (but not
reflectionally symmetric), homogeneous, incompressible MHD turbulence 
\begin{eqnarray}
    \label{eq:mean-field alpha}
    \alpha  &=&\phn\frac{\tau}{3} \left(\alpha_k + \alpha_m \right), \\
    \label{eq:mean-field alpha_k}
    \alpha_k &=& -\vec{v'} \cdot \left(\vec{\del} \cross \vec{v'} \right), \\
    \label{eq:mean-field alpha_m}
    \alpha_m &=& \phn\frac{1}{4\pi\rho} \vec{B'} \cdot \left(\vec{\del} \cross \vec{B'} \right),
\end{eqnarray}
as discussed in \cite{Pouquet_et_al_1976} and \cite{Brandenburg&Subramanian_2005}.
Here $\tau$ is the lifetime or correlation time of a typical turbulent
eddy.  In mean-field theory, these fluctuating helicities are
typically not solved directly and are instead solved through auxiliary
equations for the total magnetic helicity or are prescribed.
Here we can directly measure our fluctuating helicities and examine
whether they approximate our fluctuating emf.

To assess the possible role of an $\alpha$-effect in our
simulation, we show in Figures~\ref{fig:mean_field_alpha}$a,b$ the
fluctuating kinetic and current helicities $\alpha_k$ and $\alpha_m$ 
realized in our case~D3, averaged over the same 450~day analysis
interval. To make an estimate of the $\alpha$-effect,
we approximate the correlation time $\tau$ by defining
\begin{equation}
  \tau = \frac{H_P}{v'}
\end{equation}
where $H_P$ is the local pressure scale height and $v'$ is the local
fluctuating rms velocity, which are functions of radius only.  
Estimated by this method, the turnover time $\tau$ has a smooth radial
profile and is roughly 10~days near the bottom of the convection zone,
3~days at mid-convection zone, and slightly less near the upper boundary.
If we use the fast peak upflow or downflow velocities instead of the
rms velocities, our estimate of $\tau$ is about a factor of 4 smaller.
Our mean-field $\alpha$ (eq.~\ref{eq:mean-field alpha})
is shown in Figure~\ref{fig:mean_field_alpha}$c$.   In the upper convection zone, this is
dominated by the fluctuating kinetic helicity while the fluctuating
magnetic (current) helicity becomes important at depth.

\begin{figure}
  \includegraphics[width=9cm]{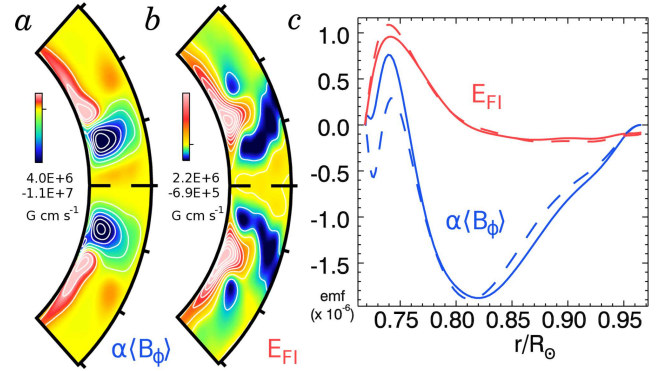}
  \caption{Comparison of emfs in case~D3.  
    $(a)$~Profile of proposed mean-field emf given by $\alpha \langle B_\phi \rangle$.
    $(b)$~Actual turbulent emf $E_\mathrm{FI}$ measured in the dynamo.
    $(c)$~Variation of hemisphere-averaged emfs with fractional
    radius.  The mean-field approximated emf is shown in blue, and
    $E_\mathrm{FI}$ in red.  The average over the northern
    hemisphere is shown solid, the southern is dashed.  
  \label{fig:emf_comparison}}
  \vspace{0.2cm}
\end{figure}

We form a mean-field emf (right-hand side of eq.~\ref{eq:mean field emf}) by
multiplying our derived $\alpha$ (Fig.~\ref{fig:mean_field_alpha}$c$) with our 
$\langle B_\phi \rangle$ (Fig.~\ref{fig:D3_bphi_production}$a$), 
and show this in Figure~\ref{fig:emf_comparison}$a$.  The turbulent emf $E_\mathrm{FI}$,
which is the left-hand side of  equation~(\ref{eq:mean field emf}),
can be measured in our simulations and is shown again in Figure~\ref{fig:emf_comparison}$b$. 
Although there is some correspondence in the two patterns, there are
significant differences.  In particular, the mean-field emf $\alpha
\langle B_\phi \rangle$ has peak
amplitudes in the cores of the wreaths (at $\pm 15^\circ$ latitude)
and is negative there.  In contrast, the actual fluctuating
emf given by $E_\mathrm{FI}$ is positive and has its highest
amplitude at the poleward side of the wreaths (near $\pm 20^\circ$
latitude).  
Thus the mean-field
emf predicts an incorrect balance in the generation
terms and would yield a distinctly different mean poloidal magnetic field. 

To assess whether better agreement may be achieved with a
latitude-averaged emf, we average the mean-field emf and $E_\mathrm{FI}$ 
separately over the northern and southern hemispheres and plot these quantities
in Figure~\ref{fig:emf_comparison}$c$.  Though both have a similar positive sense near the base of
the convection zone, the hemisphere-averaged $E_\mathrm{FI}$ becomes
small above $0.8R_\odot$ whereas the averaged mean-field emf $\alpha
\langle B_\phi \rangle$ is large and negative there.  Thus even the averaged
emfs are not in accord.

In summary, it is evident that a simple scalar $\alpha$-effect will
predict the wrong sign for the fluctuating emf in the two hemispheres,
as $\langle B_\phi \rangle$ is anti-symmetric across the equator while
$\langle A_\phi \rangle$ is symmetric.  An $\alpha$-effect based on
the kinetic helicity and magnetic helicity may capture some sense of
the fluctuating emf, as those quantities are themselves anti-symmetric
across the equator.  Yet Figure~\ref{fig:emf_comparison} suggests that
there are significant discrepancies between this particular
approximation and our turbulent emf.  In particular, this mean-field
$\alpha$-effect misses the offset between the generation regions for
mean toroidal and mean poloidal field.  This offset in latitude of the
generation regions may be important for avoiding the
$\alpha$-quenching problems encountered in many mean-field theories.
A more complex mean-field model, which takes spatial gradients of
$\langle B_\phi \rangle$ into account, may do better.  In particular,
the $\vec{\Omega} \cross \vec{J}$-effect
\citep[e.g.,][]{Moffatt&Proctor_1982, Rogachevskii&Kleeorin_2003} may
be at work in these systems, and preliminary explorations indicate
that this term matches the spatial structure of our $E_\mathrm{FI}$
better than the above $\alpha$-effect.  A tensor representation of the
$\alpha$-effect may also do much better at approximating
$E_\mathrm{FI}$, and test-field techniques could be employed to
measure this quantity \citetext{e.g., \citealp{Schrinner_et_al_2005},
and recently reviewed in \citealp{Brandenburg_2009}}.  As with our
analysis of dynamo production terms presented in \S\ref{sec:dynamo_production},
this comparative study of $\alpha \langle B_\phi \rangle$ and
$E_\mathrm{FI}$ is conducted here for the special circumstances of a
dynamo which builds global-scale magnetic fields that are nearly
steady in time.  The magnetic wreaths realized in dynamos at higher
magnetic Reynolds numbers show larger time-variations, and it is
possible that $\alpha \langle B_\phi \rangle$ better approximates
$E_\mathrm{FI}$ during the growing phase of each oscillation, when the
magnetic fields have not yet saturated in strength and the dynamo is
in a more kinematic regime.

\section{Conclusions}
\label{sec:conclusions}

The ability for a dynamo to build wreaths of strong magnetic fields in the bulk of
the convection zone has largely been a surprise, for it had generally
been supposed that turbulent convection would disrupt such magnetic
structures.  To avoid these difficulties, many solar and stellar
dynamo theories shift the burden of magnetic storage, amplification
and organization to a tachocline of shear and penetration at the base
of the convection zone where motions are more quiescent.  In contrast,
our simulations of rapidly rotating stars are able to achieve
sustained global-scale dynamo action within the convection zone
itself, with the magnetic structures both being built and able to survive
while embedded deep within the turbulence.  
These dynamos are able to circumvent the Parker instability by means
of turbulent Reynolds and Maxwell stresses that contribute to the
mechanical force balance and prevent the wreaths from buoyantly
escaping the convection zone.  
This striking behavior may be enabled by the stars rotating somewhat 
faster than the current Sun, which yields a strong
differential rotation that is a key element in the dynamo behavior.  
In our broader exploration of rapidly rotating dynamos, we find that
magnetic wreaths are present in all simulations, including those
rotating as slowly as $1.5\thinspace\Omega_\odot$.  Such structures
may  be obtainable in simulations rotating at the solar rate as well,
and efforts are underway to explore the presence of wreaths
in solar dynamos.  
  
We have achieved some dynamo states that are persistent and others
that flip the sense of their magnetic fields.  In our case~D3 the global-scale
fields have small vacillations in their amplitudes, but the magnetic
wreaths retain their identities for many thousands of days.  This
represents hundreds of rotation periods and several magnetic diffusion
times, indicating that the dynamo has achieved a persistent
equilibrium.  

Increasing the rotation rate or decreasing the magnetic diffusivity
$\eta$ yields more complex time dependence.  In many of our dynamos
the oscillations can become large, and this may result in the
global-scale fields repeatedly flipping their polarity.  At times
those dynamos appear to be cyclic but in other intervals they behave
more chaotically. Such time-dependent dynamos will be reported on in a
forthcoming paper.  In separate explorations, we have found that
magnetic wreaths also survive in the presence of a tachocline of
penetration and shear.  In those simulations the wreaths continue to
fill the convection zone even while developing roots in the
tachocline.  Dynamos in rapidly rotating suns with tachoclines can
also exhibit time-dependent oscillations and polarity reversals.
Wreath-building dynamos with tachoclines will be reported on
subsequently.

In our persistent case~D3 we are able to analyze the generation and
transport of mean magnetic field.  We find that our dynamo action is
of an $\alpha-\Omega$ nature, with the mean toroidal fields being
generated by an $\Omega$-effect from the mean shearing flow of
differential rotation.  This generation is balanced by a combination
of turbulent induction and ohmic diffusion.  The mean poloidal fields
appear to be generated by an $\alpha$-effect arising from couplings between the
fluctuating flows and fluctuating fields, with this production largely
balanced by the ohmic diffusion.  This is unlike the toroidal balance,
for here the mean flows play almost no role and the turbulent
correlations are constructive rather than destructive.  In assessing
what a mean-field model might predict for the magnetic structures
realized in case~D3, we find that the isotropic, homogeneous $\alpha$-effect based on kinetic
and magnetic (current) helicities fails to capture the sense of our turbulent
emf.  In general, our $E_\mathrm{FI}$ is poorly represented by an $\alpha
\langle B_\phi \rangle$ that is so determined.  This comparative
analysis of $\alpha \langle B_\phi\rangle$ and $E_\mathrm{FI}$ is
performed here only for the special case of a dynamo with persistent
global-scale magnetic fields.  It is possible that these results will
differ in our dynamos that show substantial time-varying behavior.

The realization of global-scale magnetic structures in
our simulations, and their great strength relative to the fluctuating
fields, may in part be a consequence of the relatively modest degree
of turbulence attained here.  Whether such structures can be generated
and sustained amidst the far more complex flows in actual stellar
interiors is not yet clear.  If such structures are indeed realized in
stars, they may or may not survive to print
through the highly turbulent convection occurring just below the stellar
photosphere.  If they do appear at the surface, some global-scale
magnetic features may propagate toward the poles along with the bands
of angular velocity speedup.   
There are some indications in stellar observations that
global-scale toroidal magnetic fields may indeed become strong in rapidly
rotating stars \citep{Donati_et_al_2006, Petit_et_al_2008}, though
small-scale fields may still account for much of the magnetic energy
near the surface \citep{Reiners&Basri_2009}.
The global-scale poloidal fields may be more successful in surviving
the passage through the turbulent surface convection.  If they do, the
stellar magnetic field will likely have significant non-dipole components.
Thus the mean poloidal fields observed at the surface may give clues to
the presence of large wreaths of magnetism that occupy the bulk of
the convection zone.

\acknowledgements
We thank Axel Brandenburg, Geoffrey Vasil, Steve Saar and Mausumi
Dikpati for helpful conversations and advice about stellar magnetism
and dynamo action.  We thank the anonymous referee for their comments
which have tightened the focus of this paper. 
This research is supported by NASA through Heliophysics Theory
Program grants NNG05G124G and NNX08AI57G, with additional support for
Brown through the NASA GSRP program by award number
NNG05GN08H.  Browning was supported by a NSF Astronomy and Astrophysics
postdoctoral fellowship AST 05-02413, and now by research support at
CITA.  Brun was partly supported by the Programme National Soleil-Terre
of CNRS/INSU (France), and by the STARS2 grant from the
European Research Council. The simulations were carried out with NSF
PACI support of PSC, SDSC, TACC and NCSA, and by NASA HEC support at
Project Columbia.  Volume renderings used in the analysis and the
field line tracings shown were produced using VAPOR
\citep{Clyne_et_al_2007}.

%
%
%
%

\appendix

\section{Production, Destruction and Transport of Magnetic Field}

We derive diagnostic tools to evaluate the generation and transport of 
magnetic field in a magnetized and rotating turbulent convection
zone.  This derivation is in spherical coordinates, and is under the
anelastic approximation.

\subsection{Induction Equation}

In the induction equation (\ref{eq:induction}), the first term on
the right hand side represents production of magnetic field while the
second term represents its diffusion.  We first rewrite the production
term to make the contributions of shear, advection and compressible
effects more explicit as
\begin{equation}
  \nab\cross(\VV\cross\BB) = (\BB\cdot\nab)\VV-(\VV\cdot\nab)\BB-\BB(\Div\VV).
\end{equation}
Under the anelastic approximation the divergence of $\vec{v}$ can be
expressed in terms of the logarithmic derivative of the mean density because
\begin{equation}
  \Div(\rb \VV)=0=\rb(\Div \VV)+(\VV\cdot\nab)\rb, \nonumber
\end{equation}
and therefore
\begin{equation}
  \Div\VV=-v_r\frac{\p}{\p r}\ln \rb.
\end{equation}
The induction equation thus becomes
\begin{equation}\label{eq:ind2}
\frac{\p \BB}{\p t}=\underbrace{(\BB\cdot\nab)\VV}_{\mbox{shearing}}
                   -\underbrace{(\VV\cdot\nab)\BB}_{\mbox{advection}}
		   +\underbrace{v_r\BB\frac{\p}{\p r}\ln \rb}_{\mbox{compression}}
		   -\underbrace{\nab\cross(\eta\nab\cross\BB)}_{\mbox{diffusion}}
\end{equation}
As labeled, the first term represents shearing of $\BB$, the second
term advection of $\BB$, the third one compressible amplification
of $\BB$, and the last term ohmic diffusion.

\subsection{Production of Axisymmetric Magnetic Field}

To identify the processes contributing to the production of mean
(axisymmetric) field, we separate our velocities and magnetic fields
into mean and fluctuating components 
$\vec{v} = \langle \vec{v} \rangle  + \vec{v}' $ and
$\vec{B} = \langle \vec{B} \rangle  + \vec{B}' $ 
where angle brackets denote an average in longitude.  Thus 
$\langle \vec{v}' \rangle = \langle \vec{B}' \rangle = 0$ by definition.
Expanding the production term of equation~(\ref{eq:ind2}) we obtain the
mean shearing term
\begin{equation}
  \langle (\BB\cdot\nab)\VV \rangle = \left( \langle \BB \rangle \cdot\nab \right) \langle \VV \rangle 
                                    + \langle (\vec{B}'\cdot\nab)\vec{v}' \rangle,
\end{equation}
the mean advection term
\begin{equation}
  -\langle (\VV\cdot\nab)\BB \rangle = -\left( \langle \VV \rangle \cdot\nab \right) \langle \BB \rangle 
                                       -\langle (\vec{v}'\cdot\nab)\vec{B}' \rangle,
\end{equation}
and the mean compressibility term
\begin{equation}
  \langle v_r\BB\frac{\p}{\p r}\ln \rb \rangle = \left( \langle v_r \rangle \langle \BB \rangle  
                                              +  \langle v_r' \vec{B}' \rangle \right)\frac{\p}{\p r}\ln \rb.
\end{equation}
In a similar fashion, the mean diffusion term becomes
\begin{equation}
  -\langle \nab\cross(\eta\nab\cross\BB) \rangle = -\nab\cross(\eta\nab\cross \langle \BB \rangle).
\end{equation}

The axisymmetric component of the induction equation is written
symbolically as:

\begin{equation}
  \label{eq:mean induction}
  \frac{\partial \langle \vec{B} \rangle }{\partial t} = P_\mathrm{MS} + P_\mathrm{FS} 
                                      + P_\mathrm{MA} + P_\mathrm{FA} 
				      + P_\mathrm{MC} + P_\mathrm{FC}
                                      + P_\mathrm{MD}
\end{equation}
With $P_\mathrm{MS}$ representing production of field by mean shear,
     $P_\mathrm{FS}$ production by fluctuating shear,
     $P_\mathrm{MA}$ advection by mean flows, 
     $P_\mathrm{FA}$ advection by fluctuating flows, 
     $P_\mathrm{MC}$ amplification arising from the compressibility of
     mean flows, $P_\mathrm{FC}$ amplification arising from
     fluctuating compressible motions, and 
     $P_\mathrm{MD}$ ohmic diffusion of the mean fields.  In turn,
     these terms are
\begin{align}
  P_\mathrm{MS} =& \left( \langle \BB \rangle \cdot\nab \right)\langle \VV \rangle, & 
  P_\mathrm{FS} =& \langle (\vec{B}'\cdot\nab)\vec{v}' \rangle, &
  P_\mathrm{MA} =& -\left( \langle \VV \rangle \cdot\nab \right) \langle \BB \rangle,& 
  P_\mathrm{FA} =& -\langle (\vec{v}'\cdot\nab)\vec{B}' \rangle, \notag\\
  P_\mathrm{MC} =&  \left( \langle v_r \rangle \langle \BB \rangle \right)\frac{\p}{\p r}\ln \rb, & 
  P_\mathrm{FC} =&  \left( \langle v_r' \vec{B}' \rangle \right)\frac{\p}{\p r}\ln \rb, \text{and} &
  P_\mathrm{MD} =& -\nab\cross(\eta\nab\cross \langle \BB \rangle). & &
  \label{eq:mean terms}
\end{align}
We now expand each of these terms into their full representation in spherical coordinates.

\subsection{Production of Mean Longitudinal Field}

\begin{eqnarray}\label{eq:ind3pe}
 \frac{\partial \langle B_{\phi} \rangle }{\partial t} &=&
             P_\mathrm{MS} + P_\mathrm{FS}  
	   + P_\mathrm{MA} + P_\mathrm{FA} 
	   + P_\mathrm{MC} + P_\mathrm{FC}
	   + P_\mathrm{MD} \nonumber\\
\label{eq:MS phi}
P_\mathrm{MS} &=& \phn\advbm  \langle v_{\phi} \rangle  + \frac{ \langle B_{\phi} \rangle  \langle v_r \rangle  + \cot\theta  \langle B_{\phi} \rangle  \langle v_{\theta} \rangle }{r}\\
\label{eq:FS phi}
P_\mathrm{FS} &=& \phn\bigg\langle \advbf v_{\phi}' \bigg\rangle  + \frac{ \langle B_{\phi}'v_r' \rangle  + \cot\theta  \langle B_{\phi}'v_{\theta}' \rangle }{r}\\
\label{eq:MA phi}
P_\mathrm{MA} &=& -\advvm  \langle B_{\phi} \rangle  - \frac{ \langle v_{\phi} \rangle  \langle B_r \rangle  + \cot\theta  \langle v_{\phi} \rangle  \langle B_{\theta} \rangle }{r}  \\
\label{eq:FA phi}
P_\mathrm{FA} &=& -\bigg\langle \advvf B_{\phi}' \bigg\rangle  - \frac{ \langle v_{\phi}'B_r' \rangle  + \cot\theta  \langle v_{\phi}'B_{\theta}' \rangle }{r} \\
\label{eq:MC phi}
P_\mathrm{MC} &=& \phn\left(\langle v_r \rangle  \langle B_{\phi} \rangle\right)\frac{\p}{\p r}\ln \rb \qquad\qquad \qquad
\label{eq:FC phi}
P_\mathrm{FC} = \left(\langle v_r'B_{\phi}' \rangle \right)\frac{\p}{\p r}\ln \rb\\
\label{eq:MD phi}
P_\mathrm{MD} &=& \phn\eta\nabla^{2} \langle B_{\phi} \rangle -\frac{\eta  \langle B_{\phi} \rangle }{r^2\sin^2\theta}+\frac{d\eta}{dr}\left(\frac{1}{r}\frac{\p (r \langle B_{\phi} \rangle )}{\p r} \right) 
\end{eqnarray}

\subsection{Production of Mean Latitudinal Field}

\begin{eqnarray}
\frac{\partial \langle B_{\theta} \rangle }{\partial t} &=&
             P_\mathrm{MS} + P_\mathrm{FS}  
	   + P_\mathrm{MA} + P_\mathrm{FA} 
	   + P_\mathrm{MC} + P_\mathrm{FC}
	   + P_\mathrm{MD} \nonumber\\
P_\mathrm{MS} &=& \phn\advbm  \langle v_{\theta} \rangle  + \frac{ \langle B_{\theta} \rangle  \langle v_r \rangle - \cot\theta  \langle B_{\phi} \rangle  \langle v_{\phi} \rangle }{r}\\
P_\mathrm{FS} &=& \phn\bigg\langle \advbf v_{\theta}' \bigg\rangle  + \frac{ \langle B_{\theta}'v_r' \rangle -\cot\theta  \langle B_{\phi}'v_{\phi}' \rangle }{r} \\
P_\mathrm{MA} &=& -\advvm  \langle B_{\theta} \rangle  - \frac{ \langle v_{\theta} \rangle  \langle B_r \rangle -\cot\theta  \langle v_{\phi} \rangle  \langle B_{\phi} \rangle }{r} \\
P_\mathrm{FA} &=& -\bigg\langle \advvf B_{\theta}' \bigg\rangle  - \frac{ \langle v_{\theta}'B_r' \rangle - \cot\theta  \langle v_{\phi}'B_{\phi}' \rangle }{r} \\
P_\mathrm{MC} &=& \phn\left( \langle v_r \rangle  \langle B_{\theta} \rangle \right)\frac{\p}{\p r}\ln \rb  \qquad\qquad \qquad
P_\mathrm{FC} = \phn\left(\langle v_r'B_{\theta}' \rangle \right)\frac{\p}{\p r}\ln \rb \\
P_\mathrm{MD} &=&\phn\eta\nabla^{2} \langle B_{\theta} \rangle 
                 +\frac{2\eta}{r^2}\frac{\p \langle B_r \rangle }{\p \theta}
                 -\frac{\eta  \langle B_{\theta} \rangle}{r^2\sin^2\theta} 
		 + \frac{d\eta}{dr}\left(\frac{1}{r}\frac{\p (r \langle B_{\theta} \rangle )}{\p r}-\frac{1}{r}\frac{\p  \langle B_r \rangle }{\p \theta} \right)
\end{eqnarray}

\subsection{Production of Mean Radial Field}

\begin{eqnarray}
\frac{\partial \langle B_{r} \rangle }{\partial t} &=&
             P_\mathrm{MS} + P_\mathrm{FS}  
	   + P_\mathrm{MA} + P_\mathrm{FA} 
	   + P_\mathrm{MC} + P_\mathrm{FC}
	   + P_\mathrm{MD} \nonumber\\
P_\mathrm{MS} &=& \phn\advbm  \langle v_r \rangle  - \frac{ \langle B_{\theta} \rangle  \langle v_{\theta} \rangle + \langle B_{\phi} \rangle  \langle v_{\phi} \rangle }{r} \\
P_\mathrm{FS} &=& \phn\bigg\langle \advbf v_r' \bigg\rangle  - \frac{ \langle B_{\theta}'v_{\theta}' \rangle + \langle B_{\phi}'v_{\phi}' \rangle }{r}  \\
P_\mathrm{MA} &=&-\advvm  \langle B_r \rangle  + \frac{ \langle v_{\theta} \rangle  \langle B_{\theta} \rangle + \langle v_{\phi} \rangle  \langle B_{\phi} \rangle }{r}  \\
P_\mathrm{FA} &=&-\bigg\langle \advvf B_r' \bigg\rangle  + \frac{ \langle v_{\theta}'B_{\theta}' \rangle + \langle v_{\phi}'B_{\phi}' \rangle }{r}  \\
P_\mathrm{MC} &=& \phn\left( \langle v_r \rangle  \langle B_r \rangle
             \right)\frac{\p}{\p r}\ln \rb \qquad\qquad \qquad 
P_\mathrm{FC} = \left( \langle v_r'B_r' \rangle \right)\frac{\p}{\p r}\ln \rb\\
P_\mathrm{MD} &=& \phn\eta\nabla^{2} \langle B_r \rangle 
                - 2\eta\frac{ \langle B_r \rangle }{r^2}
		- \frac{2\eta}{r^2}\frac{\p \langle B_{\theta} \rangle }{\p \theta}
                - \frac{2\eta \cot\theta \langle B_{\theta} \rangle }{r^2}
\end{eqnarray}

\subsection{Maintaining the Poloidal Vector Potential}

The balances achieved in maintaining the mean poloidal magnetic field are
somewhat clearer if we consider its vector potential rather than the
fields themselves.  The mean poloidal field 
$\langle \vec{B}_\mathrm{pol} \rangle$ has a corresponding vector
potential $\langle A_\phi \rangle$, where  
\begin{equation}
  \label{eq:poloidal decomposition}
  \begin{array}{rl}
    \displaystyle
    \langle \vec{B}_\mathrm{pol} \rangle &= \langle B_r \rangle \vec{\hat{r}} +
    \langle B_\theta \rangle \vec{\hat{\theta}} = \vec{\del} \cross \langle \vec{A}\big|_\phi \rangle \\[3mm]
    \displaystyle
    &= \frac{1}{r \sin \theta} \frac{\partial}{\partial \theta}
    \left\langle A_\phi \sin \theta \right\rangle \vec{\hat{r}} - 
    \frac{1}{r} \frac{\partial}{\partial r} \left\langle r A_\phi \right\rangle \vec{\hat{\theta}} \\[3mm]
    \displaystyle
    &= \vec{\del} \cross \left\langle A_\phi \vec{\hat{\phi}} \right\rangle.
  \end{array}
\end{equation}

The other components of the poloidal vector potential disappear, as terms involving
$\partial/\partial \phi$ vanish in the azimuthally-averaged equations.  
Likewise, the $\phi$-component of the possible gauge term $\vec{\del} \lambda$ is zero by
virtue of axisymmetry.
We recast the induction equation (eq.~\ref{eq:induction}) in terms of
the poloidal vector potential by uncurling the equation once and obtain
\begin{equation}
  \label{eq:A induction}
    \frac{\partial\langle\vec{A_\phi}\rangle}{\partial t} = 
        \vec{v} \cross \vec{B}{\big|}_\phi - \eta \vec{\del} \cross
        \vec{B}{\big|}_\phi.
\end{equation}
This can then be decomposed into mean and fluctuating contributions,
and represented symbolically as 
\begin{equation}
  \label{eq:A induction symbolic}
    \frac{\partial\langle\vec{A_\phi}\rangle}{\partial t} = 
        E_\mathrm{MI} + E_\mathrm{FI} + E_\mathrm{MD},
\end{equation}
with $E_\mathrm{MI}$ representing the electromotive forces (emf)
arising from mean flows and mean fields, and related to their mean
induction.  Likewise, $E_\mathrm{FI}$ is
the emf from fluctuating flows and fields and $E_\mathrm{MD}$ is 
the emf arising from mean diffusion.  These are in turn
\begin{eqnarray}
  \label{eq:A MI phi}
  E_\mathrm{MI} &= \phn\langle \vec{v} \rangle \cross \langle \vec{B} \rangle{\big|}_\phi
                 =& \langle v_r \rangle \langle B_\theta \rangle - 
                    \langle v_\theta \rangle \langle B_r \rangle, \\
  \label{eq:A FI phi}
  E_\mathrm{FI} &= \phn\langle \vec{v'}\cross\vec{B'} \rangle{\big|}_\phi
                 =& \langle v_r' B_\theta' \rangle - 
                    \langle v_\theta' B_r' \rangle, \\
  \label{eq:A MD phi}
  E_\mathrm{MD} &= -\eta \vec{\del} \cross \langle \vec{B} \rangle{\big|}_\phi 
                 =& -\eta \frac{1}{r} \left(\frac{\partial}{\partial r} 
		    \left(r \langle B_\theta \rangle \right)
		    - \frac{\partial \langle B_r \rangle}{\partial \theta}\right) 
\end{eqnarray}

\subsection{Fluctuating (Non-Axisymmetric) Component of the Induction Equation}
Left out of this analysis is the fluctuating component of the
induction equation.  This can be derived by subtracting the mean
induction equation (\ref{eq:mean induction}) from the full induction
equation, yielding the following equation for the fluctuating fields
\begin{eqnarray}\label{eq:ind4}
\frac{\p \vec{B'}}{\p t}&=&~\, ( \langle \BB \rangle \cdot\nab)\vec{v'} + (\vec{B'}\cdot\nab) \langle \VV \rangle  +\, \vec{\EE} \nonumber \\
&~&-( \langle \VV \rangle \cdot\nab)\vec{B'} - (\vec{v'}\cdot\nab) \langle \BB \rangle  -\, \vec{\FF} \nonumber \\
&~&+( \langle v_r \rangle \vec{B'} +  v_r' \langle \BB \rangle)\frac{\p}{\p r}\ln \rb +\, \vec{\GG} \nonumber \\
&~&-\nab\cross(\eta\nab\cross \langle \vec{B'} \rangle ) 
\end{eqnarray}
where the quantities $\vec{\EE}=(\vec{B'}\cdot\nab)\vec{v'}- \langle (\vec{B'}\cdot\nab)\vec{v'} \rangle $, 
$\vec{\FF}=(\vec{v'}\cdot\nab)\vec{B'}- \langle (\vec{v'}\cdot\nab)\vec{B'} \rangle $, and 
$\vec{\GG}=(v_r'\vec{B'}- \langle v_r'\vec{B'} \rangle )\frac{\p}{\p r}\ln \rb$, represent the difference between mixed
stresses from which we subtract their axisymmetric mean. In the standard mean-field derivation, these quantities 
are siblings of the G-current involving the mean electromotive force $ \langle \VV\cross\BB \rangle $ and its 3-D equivalent 
$\VV\cross\BB$ \citep[i.e., the so called ``pain in the neck'' term,][]{Moffatt_1978}.

\bibliographystyle{apj}

\end{document}